\documentclass[journal ]{new-aiaa}
\usepackage[utf8]{inputenc}
\usepackage{textcomp}
\usepackage{xspace}
\usepackage{graphicx}
\usepackage{amsmath}
\usepackage[version=4]{mhchem}
\usepackage{siunitx}
\usepackage{subcaption}
\usepackage{longtable,tabularx}
\usepackage{placeins}
\usepackage[title]{appendix}

\usepackage{hyperref}
\ifpdf
\hypersetup{
  colorlinks=true,
  linkcolor=blue,
  citecolor=blue,
  urlcolor=blue,
  pdftitle={Wide-scale Monitoring of Satellite Lifetimes: \linebreak
Pitfalls and a Benchmark Dataset},
  pdfauthor={} 
}
\fi

\usepackage{cite}

\newcommand{\tableRef}[1]{\autoref{#1}}
\newcommand{\figRef}[1]{\autoref{#1}}

\newcommand{\secRef}[1]{\autoref{#1}}
\newcommand{\appendixRef}[1]{appendix \ref{#1}}

\newcommand{\etal}{{\em \xspace et al.}}
\newcommand{\eg}{{\em e.g., }}

\title{Wide-scale Monitoring of Satellite Lifetimes: \linebreak
Pitfalls and a Benchmark Dataset}

\author{David P. Shorten}
\affil{School of Mathematical Sciences, The University of Adelaide}

\author{Yang Yang}
\affil{School of Mechanical and Manufacturing Engineering,  The University of New South Wales}

\author{John Maclean}
\affil{School of Mathematical Sciences, The University of Adelaide}

\author{Matthew Roughan}
\affil{Teletraffic Research Centre and School of Mathematical Sciences, The University of Adelaide}

\begin{document}

\maketitle

\begin{abstract}
    An important task within the broader goal of Space Situational
    Awareness (SSA) is to observe changes in the orbits of satellites,
    where the data spans thousands of objects over long time scales
    (decades).  The Two-Line Element (TLE) data provided by the North
    American Aerospace Defense Command is the most comprehensive and
    widely-available dataset cataloguing the orbits of
    satellites. This makes it a highly-attractive data source on which
    to perform this observation. However, when attempting to infer
    changes in satellite behaviour from TLE data, there are a number
    of potential pitfalls. These mostly relate to specific features of
    the TLE data which are not always clearly documented in the data
    sources or popular software packages for manipulating them.  These
    quirks produce a particularly hazardous data type for researchers
    from adjacent disciplines (such as anomaly detection or machine
    learning). We highlight these features of TLE data and the
    resulting pitfalls in order to save future researchers from being
    trapped. A seperate, significant, issue is that existing
    contributions to manoeuvre detection from TLE data evaluate their
    algorithms on different satellites, making comparison between
    these methods difficult.  Moreover, the ground-truth in these
    datasets is often poor quality, sometimes being based on
    subjective human assessment. We therefore release and describe
    in-depth an open, curated, benchmark dataset containing TLE data
    for 15 satellites alongside high-quality ground-truth manoeuvre
    timestamps.
\end{abstract}
 
\section{Introduction}

The ability to recover the manoeuvre history of satellites is
important for the analysis of their lifetimes.  In particular, it is
of significant interest to know whether a satellite has already
reached its end of life and if its last manoeuvre has been performed
in accordance with the Inter-Agency Space Debris Coordination
Committee (IADC) mitigation guidelines~\cite{lemmens2014tle}, or to
quickly discover unplanned manoeuvres.

Large-scale and long-term historical datasets are essential for
inferring satellite manoeuvres in the post-mission mode. Detailed tracking measurements, such
as optical and radar data, are not publicly available for most
satellites, but the Two-Line Element (TLE) data~\cite{celesTrak}
provided by the North American Aerospace Defense Command (NORAD) does provide
long-term data detailing the orbits of a large number of satellites.
The TLE data contains perturbation mean orbital elements, calculated
by least squares from tracking observations of a space object
\cite{vallado2012two, vallado2006revisiting} and sampled at roughly
daily intervals (more detail on sampling will follow).  A more
accurate catalogue, known as the {Special Perturbation} (SP)
catalogue, is maintained by the $18^{th}$ Space Control Squadron and
contains the sensing data of the United States Space Surveillance
Network (SSN)\cite{scs2020spaceflight}. TLEs are increasingly an
outcome of fitting satellite ephemeris in the SP catalogue
\cite{pastor2021manoeuvre}. TLE data is publicly available from the
U.S. Space Command’s space object catalog via
Space-Track\footnote{\url{https://www.space-track.org}} or
Celestrak\footnote{\url{https://celestrak.com}}.

Most active satellites perform routine manoeuvres to change their
orbital altitude or plane. A planned satellite manoeuvre is performed
using the satellite's propulsion subsystem to fire thrusters and bring
about a change in its orbital elements, which otherwise would remain
constant. Simple time series analysis on these orbital elements can
therefore be employed to detect satellite manoeuvres. These can be
enhanced by screening on TLE prediction residuals using a physical
model to predict the future motion of the satellite. In particular TLE
data is intended to be used in conjunction with the SGP4/SDP4
model~\cite{hoots1980spacetrack}, which can perform such predictions.

As explained in depth in \secRef{sec:TLE_description}, this model
allows one to derive satellite positions across multiple time points
from a single TLE. However, in this work we show that, if one's goal
is to detect changes in a satellite's orbit over longer time periods
(days or weeks), then the use of this model can be
counter-productive. Specifically, in \secRef{sec:pitfalls:full_sgp4}
and \secRef{sec:pitfalls:mean_to_propagations}, we show that using the
full SGP4/SDP4 model to propagate to subsequent epochs and make
comparisons against the associated TLEs is not necessarily better than
performing the simpler propagation of the mean orbital elements. In
\secRef{sec:pitfalls:osculating_at_tle_timestamps}, we demonstrate how
using SGP4/SDP4 to add in non-Keplerian orbit components before
performing further analysis can transform the data in unexpected
ways. Thus, one contribution of this work is to clarify the use of TLE
data in conjunction with satellite models for change detection. That
might seem minor, but we identify many past papers whose results might
be more convincing with a better command of these important details.

This contribution can be summarised in three specific recommendations:
\begin{enumerate}
    \item If possible, avoid propagating TLEs using the full SGP4/SDP4 model.
    \secRef{sec:pitfalls:full_sgp4} shows how using this model in its entirety introduces
    substantial noise in the propagation residuals, as compared with propagating mean elements.
    \item Avoid transforming TLEs to osculating elements or satellite positions using
    SGP4/SDP4. If this is done, ensure that a sufficiently high sampling rate is used.
    \secRef{sec:pitfalls:osculating_at_tle_timestamps} demonstrates how a low sampling rate
    can lead to aliasing of the high-frequency non-Keplerian components introduced by SGP4/SDP4.
    \item If one does propagate TLEs using the full SGP4/SDP4 model, then ensure that TLEs in subsequent
    epochs are also transformed into osculating elements or satellite positions using this model.
    \secRef{sec:pitfalls:mean_to_propagations} shows the problems that can arise if this step is not included.
\end{enumerate}

The second major contribution of this work is a dataset on which to
base accurate results for future detection projects.  Much of the
previous work on detecting changes in satellite orbits from TLE data
(see \secRef{sec:previous_work} for a review) has tested techniques on
only a limited number of satellites. Moreover, each work chooses
different satellites on which to test, and the ``ground-truth'' used is
sometimes not a true ground truth as it is obtained from inspection of
the same data being used to perform detection.  In order to address
these concerns, \secRef{sec:dataset_description} presents an open,
curated dataset of satellite TLEs and associated manoeuvre timestamps
to serve as a benchmark for future work on satellite lifetime
surveillance. This dataset can be found at
\texttt{\url{github.com/dpshorten/TLE_observation_benchmark_dataset}}.
 
\section{Previous Work}
\label{sec:previous_work}

Change detection in satellite data now forms a small cottage industry,
which we shall summarise below, but some definitions are needed first:
each {\em element} of TLE data consists of two lines of data including
an associated timestamp, which is referred to as its {\em
  epoch}~\cite{CelestrakFAQ}, and the satellites orbital elements.
The epoch is the notional time at which the data is recorded, though
in fact the element may summarise a fit to data from a wider time
interval. In general, we use the term epoch to mean a distinct point
in time at which we can measure or predict a satellite position.

A popular approach to detecting changes in the orbits of satellites
from TLE data is to propagate the TLE published for one or more epochs
to the epoch associated with another TLE and compare the propagated
(or predicted) and measured values~\cite{li2018new, li2019maneuver,
  decoto2015technique, mukundan2021simplified,zhao2014method}. The
propagation should be performed using the SGP4/SDP4
model~\cite{hoots1980spacetrack,CelestrakFAQ, vallado2006revisiting,
  vallado2012two} (see \secRef{sec:TLE_description} for a detailed
explanation of the relationship between TLE data and the SGP4/SDP4
model). Some work \cite{mukundan2021simplified, zhao2014method}
utilising this general approach simply propagates a TLE to the
subsequent epoch, whereas others \cite{decoto2015technique,
  lemmens2014two, li2018new, li2019maneuver} propagate the TLEs from
multiple epochs (both forwards and backwards in time) to a given epoch
before making a comparison. As will be described in detail in
\secRef{sec:pitfalls:mean_to_propagations}, great care must be taken
to ensure that the TLE published at the epoch to which the other
epochs are being propagated is also transformed using SGP4/SDP4 in
order to have the non-Keplerian components of the orbit included. Much
of the literature \cite{li2019maneuver, mukundan2021simplified,
  li2018new, zhao2014method} does not make it clear whether or not
this step is performed. However, in some studies
\cite{mukundan2021simplified, zhao2014method}, the large differences
between the propagated orbital elements and those at subsequent epochs
imply that it is likely that the non-Keplerian terms are not being
added at the subsequent epochs.

There is also a substantial body of work which looks for changes in the TLE data using a data-centric approach, that
is, without taking into account the underlying orbital dynamics
\cite{song2012simple, agueda2013orbit, kelecy2007satellite, patera2008space, swartz2010swift, shivshankar2021behaviour,
bowman2015abnormal, kraus2012detecting, roberts2021geosynchronous, wang2022machine, wang2021gaussian}.
The dominant data-centric approach uses some method
to predict the values in a TLE based usually on past TLE values, but sometimes also including future TLEs in the
predictions. If the recorded TLE deviates substantially from the predictions, then this is taken to be indicative of
some change in the satellite's orbit. The simplest such approach is that published by
Song\etal\cite{song2012simple}, who  applied outlier detection methods to changes in the semi-major axis between
subsequent epochs. A number of studies applied polynomial regression to the prediction of TLE updates
\cite{agueda2013orbit, kelecy2007satellite, patera2008space, swartz2010swift}.
Shivshankar and Ghose \cite{shivshankar2021behaviour} apply classic time-series forecasting techniques to this problem.
Two studies \cite{bowman2015abnormal, kraus2012detecting} used neural networks to predict TLE values.

An alternative data-centric approach focuses specifically on the
problem of detecting the timestamps of satellite manoeuvres and solves
this task using supervised machine learning \cite{wang2022machine,
  wang2021gaussian}. That is, these studies compile datasets of TLE
data and associated ground-truth manoeuvre timestamps and train models
on this combination in order to be able to predict the presence of
manoeuvres in unseen data.  The study by
Wang\etal\cite{wang2021gaussian} is unusual here, as, unlike other
studies which make use of TLEs published by NORAD \cite{CelestrakFAQ},
they only make use of simulated data.

While the above-mentioned data-centric approaches operated on the raw TLE values (or a simple transformation of them,
such as transforming the mean motion to the semi-major axis),
Roberts and Linares \cite{roberts2021geosynchronous} first created a regularly-sampled time series by propagating the
TLEs using SGP4/SDP4 to sample points.
A convolutional neural network is trained on a set of this data for which labelled manouevre timestamps are also provided.

An interesting counterpoint to these studies using supervised machine learning is the work of
Shabarekh\etal\cite{shabarekh2016efficient, shabarekh2016novel}, who present machine learning approaches for
predicting satellite manoeuvre timestamps based solely on historic manoeuvre times (that is, without any TLE data).

There are two notable studies \cite{bai2019mining, lemmens2014two} that develop both data-centric and
propagation-based approaches and contrast the two.

Finally, there are two significant recent studies that focus on the problem of inferring manoeuvre parameters from the
TLEs at the epochs on either side of a given manoeuvre.
Yu\etal\cite{yu2021maneuver} developed a model-based approach for determining the time point of a manoeuvre based
on the mean orbital elements before and after the manoeuvre. It depends on preexisting knowledge that the manoeuvre
took place.
Pirovano and Armellin \cite{pirovano2022detection}, develop an optimisation framework for estimating
manoeuvre parameters for an unknown manoeuvre occurring between two TLE epochs. They show how this approach can also
be adapted to manoeuvre detection.

The striking factor in these studies, particularly the data-centric
studies, is that TLE data is often used in a manner that does not
display a clear understanding on the non-Keplerian elements of orbital
dynamics, and that can cause problems as we show below. It is towards
this gap that this paper and the associated dataset are aimed. 
 
\section{Description of TLE Data}
\label{sec:TLE_description}

TLE is a data format encoding the orbital elements of Earth-orbiting
objects. The data is encoded in ASCII text files as a sequence of
characters and numbers. A single {\em element} is actually encoded over
three lines in the form:

\begin{verbatim}
ISS (ZARYA)
1 25544U 98067A   08264.51782528 -.00002182  00000-0 -11606-4 0  2927
2 25544  51.6416 247.4627 0006703 130.5360 325.0288 15.72125391563537
\end{verbatim}

\noindent where the first line encodes the 24 character name of the
object, and the first data line contains information about the
satellite (classification, international designator, ...)  and the
epoch (the time point to which the measurement pertains). The second
data line contains the orbital elements. Both lines have an additional
checksum. The data are formatted over multiple lines because the
format's original use was on 80-column punch cards.

The pitfalls that can occur when analysing TLE data stem from some misunderstandings around the
 nature of what this data represents. This is understandable, as certain features of this data are often not
clearly documented. This lack of clear documentation can make it difficult for those from other domain
areas (such as anomaly detection or machine learning) to apply their knowledge to this area. As such, we first
describe in some detail what the values in TLE data do represent, with a focus on making concepts clear to
researchers unfamiliar with satellite orbit data. Subsequent sections will focus on some of the pitfalls that
these misunderstandings can lead to.

The main potential source of confusion around TLE data is that reported numerical
values are quite far removed from the underlying physical quantities. Anomalies in the data can therefore be the
result of changes in how the underlying physical quantities are being transformed, as opposed to changes in the
quantities themselves. This is in contrast with, say, financial market data, where the price associated with
a stock bid is closely associated with the actual stock-market behaviour. Further, the relationship between TLE
data and the models for propagating it is not straightforward (see below).

In an idealised model of an orbit, we would consider a small point
mass (a satellite) orbiting a substantially larger point mass (the
earth). This will result in an elliptical orbit which can be described
by five parameters, typically the parameterisation used is: the
semi-major axis ($a$), eccentricity ($e$), inclination ($i$),
longitude of the ascending node ($\Omega$) and argument of the
periapsis ($\omega$). A sixth parameter, the mean anomaly ($M$),
describes the point along this ellipse where the satellite sits at a
given epoch. These six parameters are usually referred to as Keplerian
elements \cite{schaub2003analytical, vallado2001fundamentals}.

\begin{figure}[ht!]
       \centering
       \includegraphics[width=\linewidth]{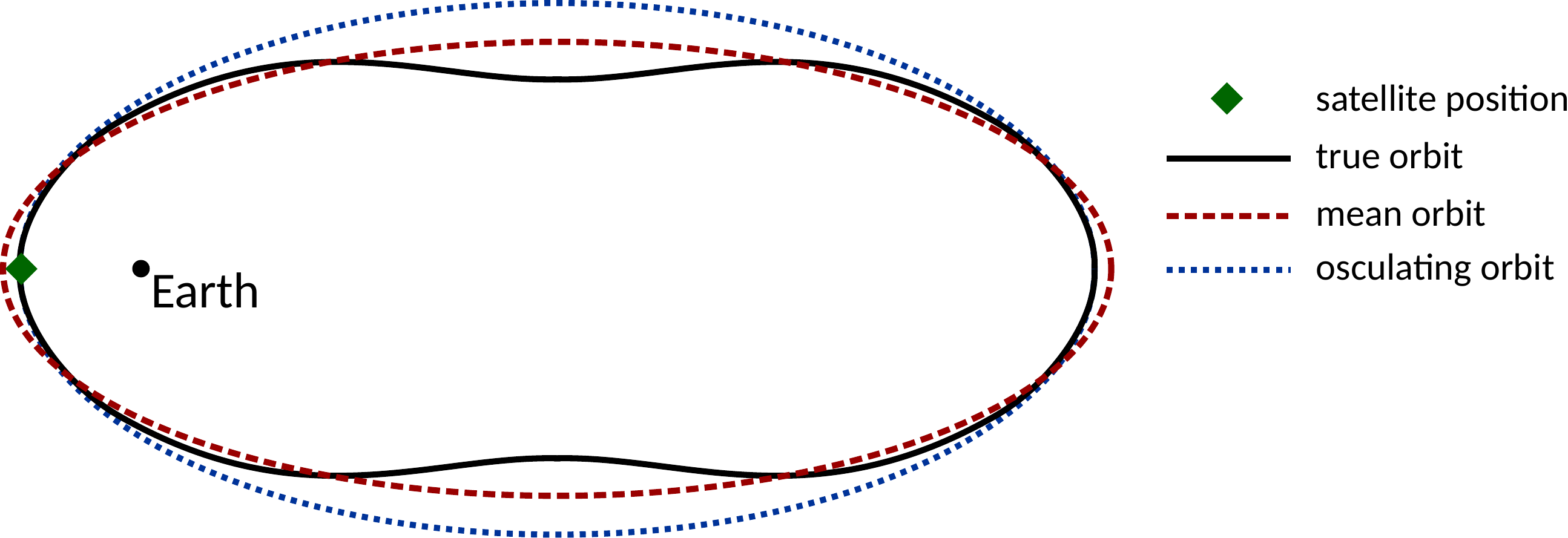}
       \caption{An illustration of mean and osculating orbits. The
         figure shows a satellite and its true orbit (which is
         non-elliptical due to perturbations), the mean (elliptical)
         orbit best fitting its overall path, and the osculating
         (elliptical) orbit for the satellite at perigee. Note that
         all affects are strongly exaggerated to make the illustrated
         differences more obvious. In particular one can see that
         using the mean orbit to estimate position will only be valid
           on a set of measure zero, but likewise, propagating the
         osculating orbit will be inaccurate at any point except the
         one for which it is calculated.}
       \label{fig:osculating_v_mean}
\end{figure}

There are other parametrisations which can describe an orbit. Notable in the current context would be the
three-dimensional position and velocity vectors $\vec{x}$ and $\vec{v}$ \cite{schaub2003analytical, vallado2001fundamentals}.
It is easy to transform these vectors into the Keplerian elements using the mass of the larger object.

Real satellites, however, do not orbit in perfect ellipses
\cite{vallado2001fundamentals}. This is due to factors such as the
gravitational pull of other bodies (such as the moon) and atmospheric
drag. A question then arises: given that the orbit is not elliptical,
if we are given position and velocity vectors, how do we transform
these into Keplerian orbital elements? There are two Keplerian orbits
that can be used to describe the motion of the satellite: the $mean$
and $osculating$ orbits \cite{schaub2003analytical,
  vallado2001fundamentals}.

\autoref{fig:osculating_v_mean} illustrates the two elliptical
representations of an orbit, though note it is substantially
exaggerated. The osculating orbit is the Keplerian orbit that
``kisses'' the true orbit at the point at which the satellite
currently sits. That is, it is the Keplerian orbit which will most
closely approximate the motion of the satellite in the small
neighbourhood around its current position. By contrast, the mean orbit
is the Keplerian orbit which will most closely approximate the
satellite's orbit across its entire orbital period. This distinction
is important when we are transforming Keplerian orbital elements into
position vectors. If we translate the mean Keplerian orbital elements
into position and velocity vectors in the straightforward manner that
we might do for an elliptical orbit \cite{bate2020fundamentals}, then
these vectors might differ significantly from the actual position and
velocity.

Most importantly for change detection, we note that the mean orbital
elements remain approximately constant, but inaccurate, while the
osculating elements are accurate, but depend on the satellite
location, so they vary periodically. 

TLE data contains the {\bf mean Keplerian orbital elements} (apart
from the semi-major axis, although this can easily be recovered from
the mean motion) \cite{hoots1980spacetrack, vallado2006revisiting}.
TLE data is intended to be used in conjunction with the SGP4/SDP4
orbital propagator. Although it is often referred to as a
`propagator', the purpose of SGP4 is not only to propagate orbits to
later points in time, but also to recover the osculating orbit (and
precise satellite positions and velocities) from the mean orbit
\cite{hoots1980spacetrack, vallado2006revisiting,
  zhao2014method}. Doing so also requires the satellite's ballistic
drag coefficient, which is provided in the TLE data. The implication
of this is that the orbital elements contained in the TLE data can
give a poor estimation of the current position and velocity of the
satellite. It is only when the non-Keplerian components of the orbit
are reintroduced via the SGP4/SDP4 model that we arrive at a reliable
estimation of the satellite's current state.

\begin{figure}[ht!]
       \centering
       \includegraphics[width=\linewidth]{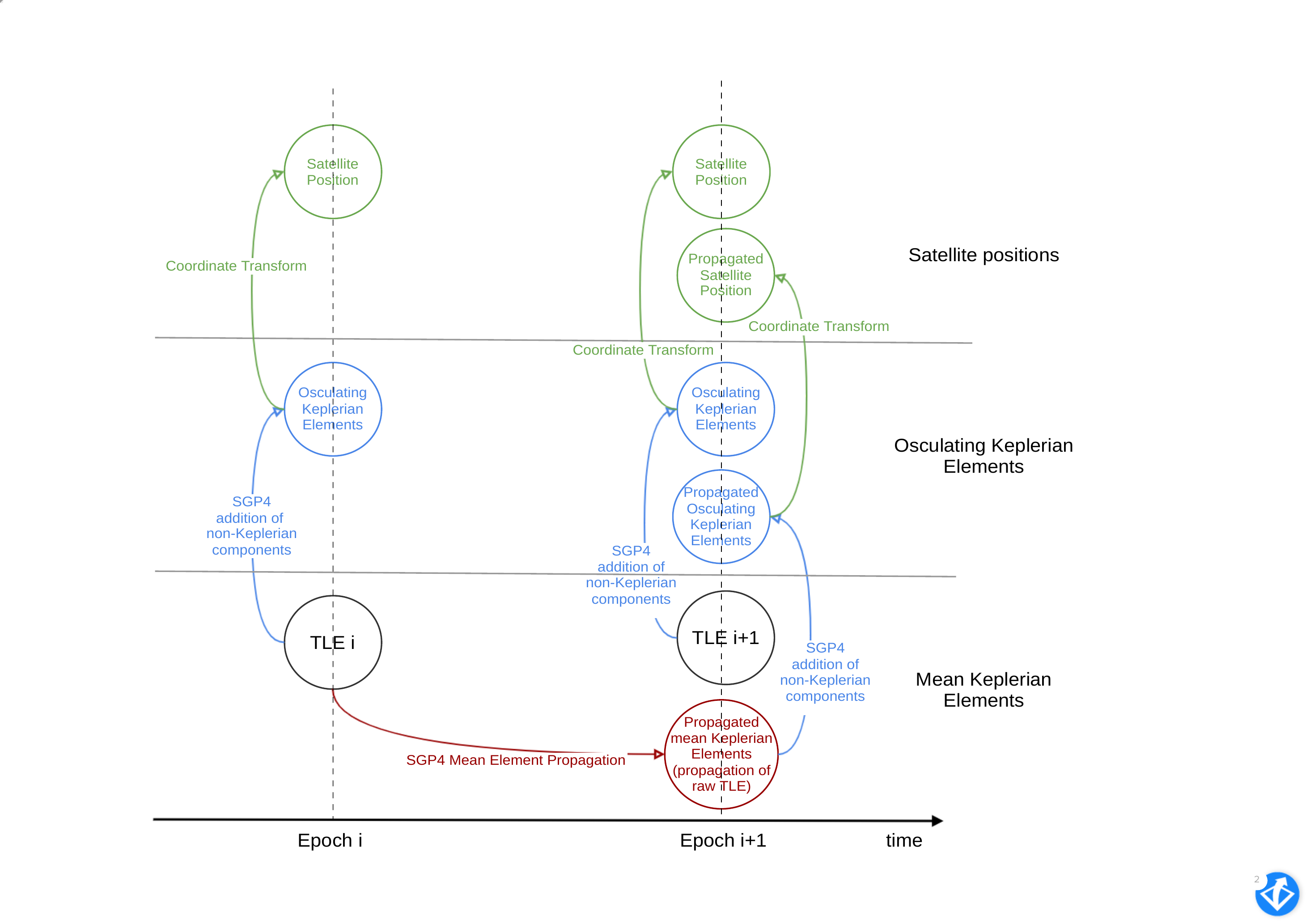}
       \caption{The process by which a TLE record can be used and
         propagated to a subsequent epoch. The goal of propagation is
         often to compare the propagated values to measured values at
         a subsequent epoch. There are a number of ways that this
         comparison can be done. One could compare the mean Keplerian
         elements propagated by SGP4/SDP4 with the mean orbital
         elements in the TLE recorded at the subsequent
         epoch. Alternatively, one could use SGP4/SDP4 to incorporate
         the non-Keplerian components of the orbit in order to
         generate a precise position for the satellite at the instant
         of the TLE epoch. This could be expressed as either
         osculating Keplerian elements or as position and velocity
         vectors.}
       \label{fig:TLE_description:mean_osculating_explanation_diagram}
\end{figure}

Popular software libraries for manipulating TLE data (such as \texttt{Skyfield}
\cite{python-skyfield} and \texttt{pyorbital} \cite{pyorbital}) obscure some of these details for ease of use. These
libraries make it easy to load TLE data and then query satellite positions or orbital elements.
However, if we query the orbital elements or satellite positions at the same time point
as a given TLE epoch, we will not be returned the original orbital elements of the TLE or a simple conversion from
these elements to positions. Rather, the SGP4/SDP4 model is used to incorporate the non-Keplerian terms not present in
the TLE data before returning any values to the user.

We need to take great care in how we treat orbital elements or positions to which these non-Keplerian components have
been added. Some of these non-Keplerian components have periods much shorter than the interval between
published TLE datapoints. This means that these components can be
highly {\em aliased} \cite{marks2009handbook}.
Moreover, NORAD can and does make changes in the relative timepoints at which TLEs
are published, which will change the effect of this aliasing. This phenomenon is demonstrated on a real satellite in
\secRef{sec:pitfalls:osculating_at_tle_timestamps}.

The SGP4/SDP4 model first propagates the mean orbital elements, before the non-Keplerian terms  are incorporated (see
page 12 of \cite{hoots1980spacetrack}). It is possible to use this first component
separately to model the evolution of the mean elements. Indeed, the very popular \texttt{python-sgp4} library
\cite{python-sgp4} (used by \texttt{Skyfield}) offers this functionality.
\figRef{fig:TLE_description:mean_osculating_explanation_diagram}
provides a diagram showing the various modifications that can be made to TLEs, particularly within the context of
comparing the propagation of one TLE to the TLE at the subsequent epoch. There are two ways
of doing this. We could propagate the first TLE to the subsequent epoch using the full SGP4/SDP4 model. We would then
also
need to add the non-Keplerian terms to the TLE at the subsequent epoch with SGP4/SDP4. We could then compare the
resulting osculating orbital elements or satellite positions. An alternative approach would be to just propagate the
mean elements from the first TLE and compare these propagated elements directly against the values recorded in the
subsequent TLE. \secRef{sec:pitfalls:full_sgp4} contrasts these two approaches in more depth. It is shown that the
use of the full SGP4/SDP4 model can be detrimental to the purpose of detecting long-term changes in satellite orbits.
 
\section{Potential Pitfalls}
\label{sec:pitfalls}

In this section, we analyse three pitfalls that can be encountered when analysing TLE data for
the surveillance of satellite lifetimes. The first pitfall (\secRef{sec:pitfalls:full_sgp4}) occurs with
regularity in published research \cite{bai2019mining, decoto2015technique, zhao2014method, lemmens2014two} and
also snared the authors in preliminary research using TLE data. The second pitfall
(\secRef{sec:pitfalls:osculating_at_tle_timestamps}) is one that the authors have personally encountered and also
occurs at least once \cite{roberts2021geosynchronous} in the published literature. It is difficult to confirm whether
 the third pitfall (\secRef{sec:pitfalls:mean_to_propagations}) was encountered in a given study without
inspecting the scripts used to run the analysis. However, there are studies
\cite{mukundan2021simplified, li2018new, li2019maneuver} where the figures suggest that it
might have played a role.

\subsection{Pitfall 1: Including non-Keplerian Components in Propagation}
\label{sec:pitfalls:full_sgp4}

This subsection makes the recommendation that, if one propagates TLEs to subsequent
epochs in order to make comparisons with the TLEs recorded at those epochs, then it
is preferable to only propagate the mean elements. This is as opposed to using the
full SGP4/SDP4 model.

The most common approach towards detecting anomalous updates in TLE
data is to propagate the TLE from a given epoch to subsequent epochs
and to then compare the propagated positions with those recorded from
those in subsequent epochs \cite{bai2019mining, decoto2015technique,
  zhao2014method, lemmens2014two, mukundan2021simplified, li2018new,
  li2019maneuver}.  There are also more complicated variations on this
theme, such as propagating over multiple epochs or propagating
backwards in time, or compositions of methods into larger schemes.

To the best of the authors' knowledge, existing research that utilises
this approach performs the propagation using the full SGP4/SDP4 model,
which includes the non-Keplerian components.  As described in detail
in \secRef{sec:TLE_description}, the orbital elements in the TLEs
correspond to the mean orbit of the satellite, which is the Keplerian
orbit which best approximates its behaviour throughout a full orbit.
Finding the position vectors which correspond with these mean
Keplerian elements does not make sense, as the resulting positions
will correspond to points on a fictional orbit, as opposed to the
actual satellite positions. The standard practice for obtaining
satellite positions from TLE records is, therefore, to use the
SGP4/SDP4 model to reintroduce the non-Keplerian orbital components
before calculating positions \cite{vallado2006revisiting}.
Comparisons to subsequent (or past) records are then usually made
based on the satellite positions.  Most often, comparison is made in
the radial, in-track and cross-track directions, as these correspond
to the directions in which thrust is applied during manoeuvres
\cite{milani1987non}.  Given that this is the standard approach, and
that current software implements this approach by default, we make the
assumption that in previous work when comparisons of position vectors
are made, unless explicitly stated, these positions are arrived at by
using the full SGP4/SDP4 model.

This subsection argues that using the full SGP4/SDP4 model for propagation
\cite{li2018new, li2019maneuver, decoto2015technique, mukundan2021simplified, zhao2014method}
introduces unnecessary complexity into the analysis. The SGP4/SDP4 model first propagates the mean orbital elements,
before incorporating the non-Keplerian terms (see page 12 of \cite{hoots1980spacetrack}). A
simpler analysis technique propagates the mean elements to the desired epoch and then
compares the propagated mean elements directly against the raw TLE elements (a straightforward transform between the
mean motion and semi-major axis is required). This approach greatly reduces the complexity of the comparison process.
Only one of the two TLEs needs to be modified by SGP4/SDP4. Moreover, this modification only involves a subset of the
usual terms of this model.

\begin{figure}[ht!]
       \centering
       \begin{subfigure}[]{0.6\linewidth}
           \includegraphics[width=\linewidth]{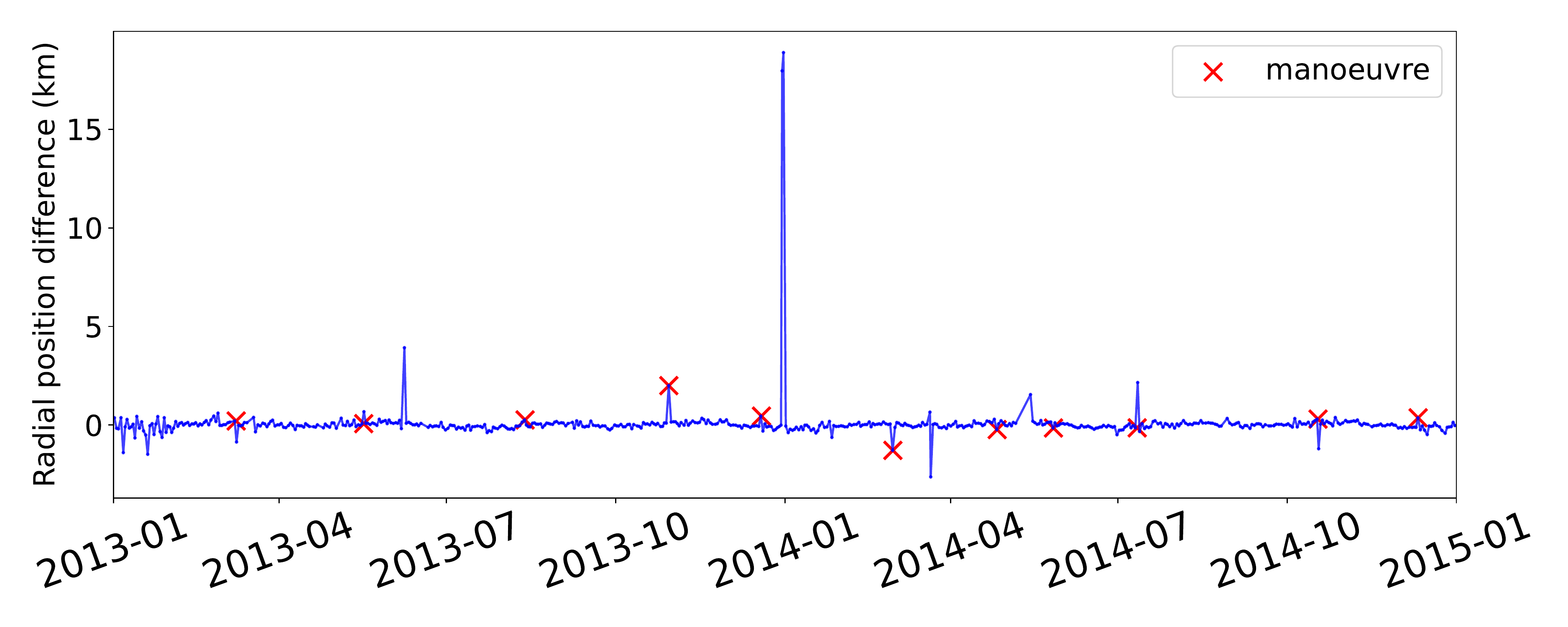}
           \caption{Radial position difference, one epoch propagation}
           \label{fig:pitfalls:propagation_type_comparison:radial}
       \end{subfigure}
       \begin{subfigure}[]{0.45\linewidth}
           \includegraphics[width=\linewidth]{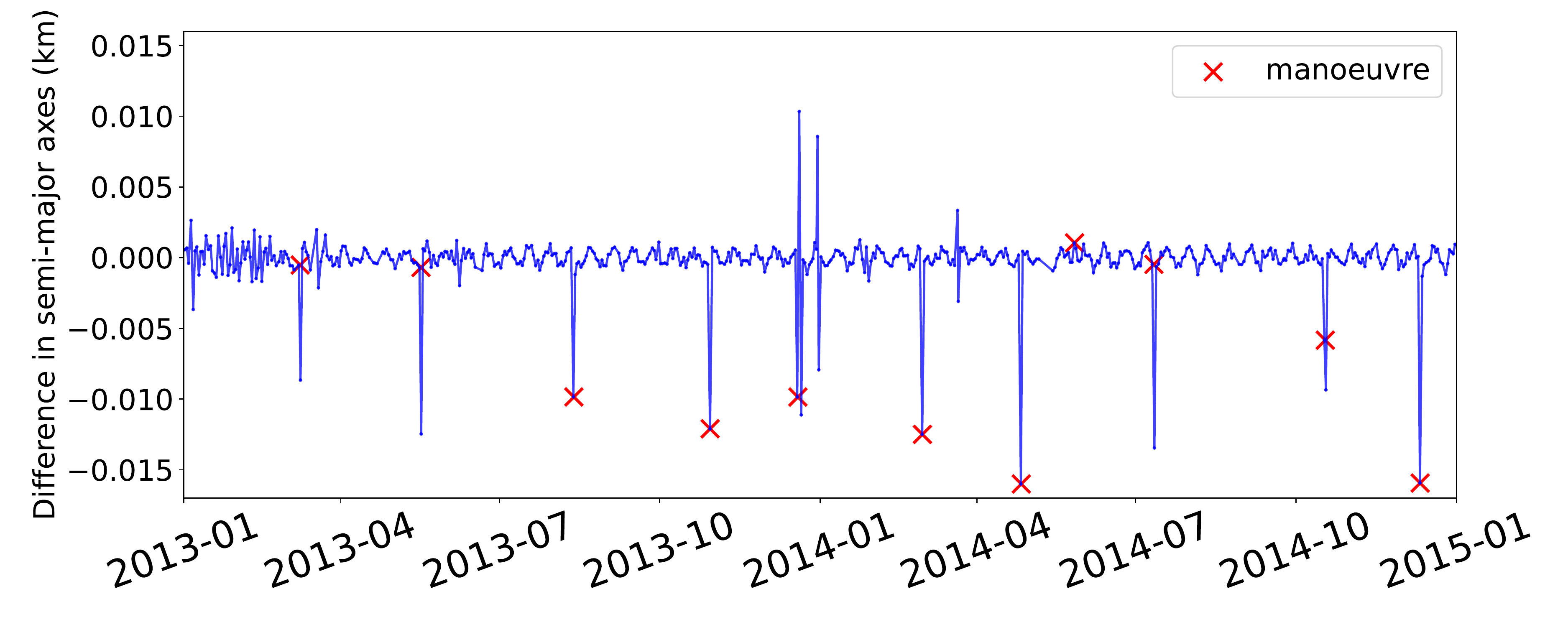}
           \caption{Difference in mean semi-major axes, one epoch propagation}
           \label{fig:pitfalls:propagation_type_comparison:mean_a}
       \end{subfigure}
       \begin{subfigure}[]{0.45\linewidth}
           \includegraphics[width=\linewidth]{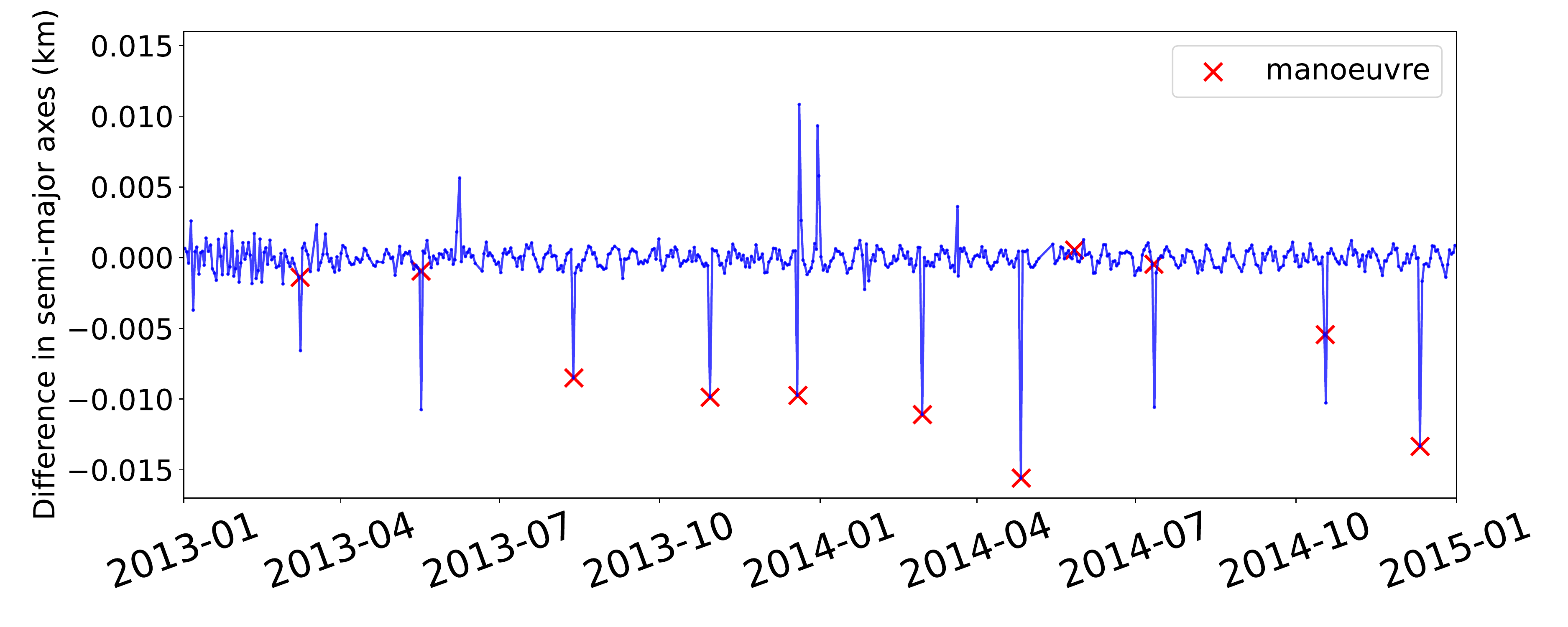}
           \caption{Difference in osculating semi-major axes, one epoch propagation}
           \label{fig:pitfalls:propagation_type_comparison:osculating_a}
       \end{subfigure}
       \begin{subfigure}[]{0.45\linewidth}
           \includegraphics[width=\linewidth]{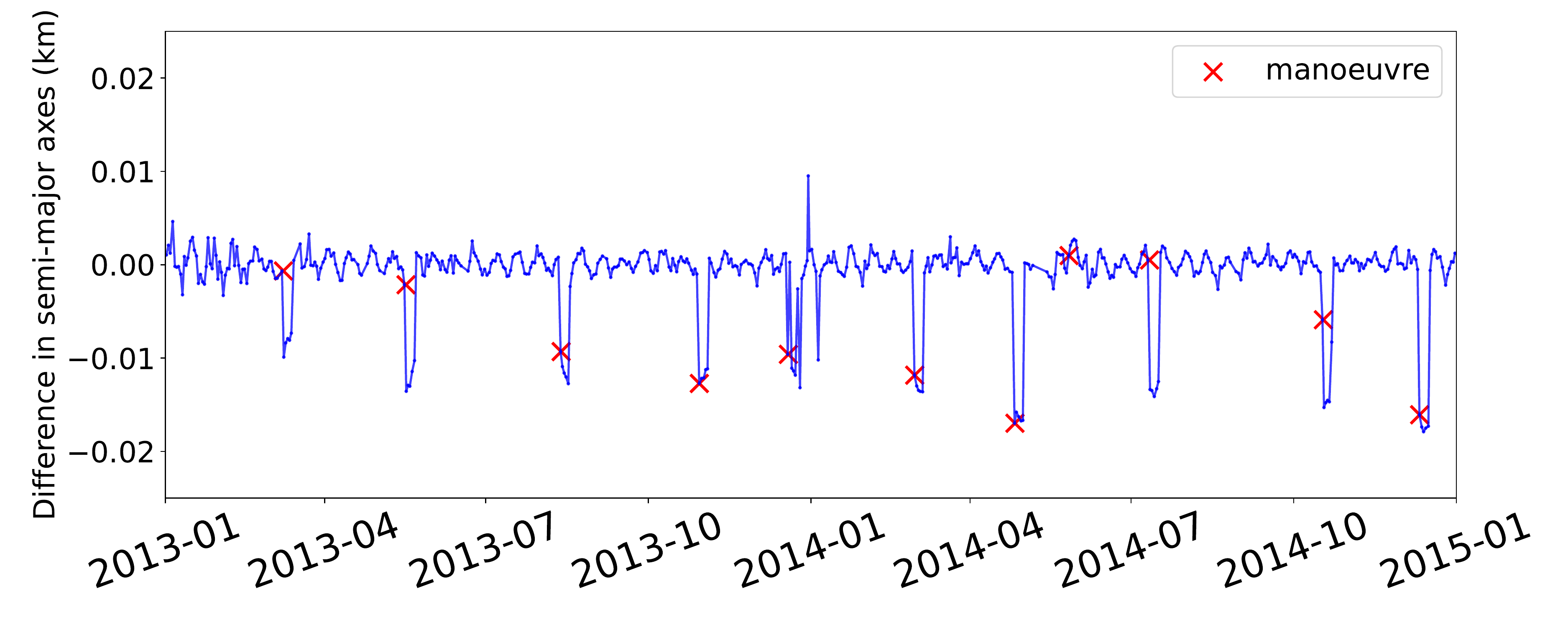}
           \caption{Difference in mean semi-major axes, five epoch propagation}
           \label{fig:pitfalls:propagation_type_comparison:mean_a_offset_5}
       \end{subfigure}
       \begin{subfigure}[]{0.45\linewidth}
           \includegraphics[width=\linewidth]{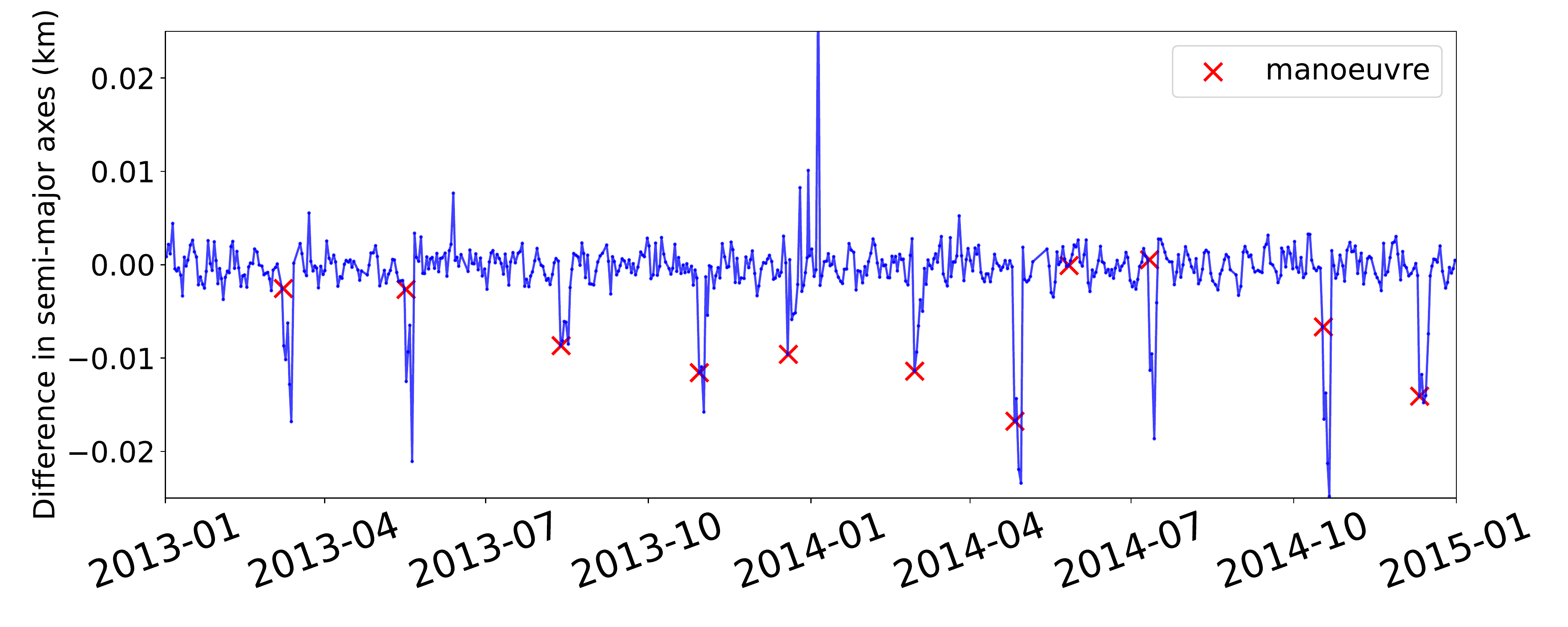}
           \caption{Difference in osculating semi-major axes, five epoch propagation}
           \label{fig:pitfalls:propagation_type_comparison:osculating_a_offset_5}
       \end{subfigure}
       \caption{ \small The different techniques for propagating
         a TLE to a subsequent epoch and then comparing the
         propagation to the TLE associated with that epoch for the
         Jason 2 satellite in the years 2013 and 2014. Independently
         obtained manoeuvre timestamps are shown with red crosses.
         Subfigure~(\protect\subref{fig:pitfalls:propagation_type_comparison:radial})
         shows the result of performing propagation using the full
         SGP4/SDP4 model and then computing differences in the radial
         position. Note that some large propagation differences (\eg
         around Jan 2014) do not align well with manoeuvre
         timestamps. \figRef{fig:pitfalls:high_frequency_positions} in
         \appendixRef{sec:appendix:additional_pitfall_1} shows similar
         phenomena in both the in-track and cross-track
         directions. Subfigures~(\protect\subref{fig:pitfalls:propagation_type_comparison:mean_a})
         and
         (\protect\subref{fig:pitfalls:propagation_type_comparison:osculating_a})
         show differences between propagated and measured values for the semi-major axis. The
         magnitude of these differences corresponds fairly well with the
         manoeuvre timestamps.
         Subfigure~(\protect\subref{fig:pitfalls:propagation_type_comparison:mean_a})
         shows the difference between propagating the mean semi-major
         axis and the mean semi-major axis in the subsequent TLE
         (obtained from the mean motion), whereas
         (\protect\subref{fig:pitfalls:propagation_type_comparison:osculating_a}),
         By contrast, propagates the semi-major axis using the full
         SGP4/SDP4 model and compares the propagations against the
         osculating element in the subsequent epoch, also obtained
         using SGP4/SDP4. There are small, but significant,
         differences in the propagation errors shown in
         (\protect\subref{fig:pitfalls:propagation_type_comparison:mean_a})
         and
         (\protect\subref{fig:pitfalls:propagation_type_comparison:osculating_a}).
         Subfigures~(\protect\subref{fig:pitfalls:propagation_type_comparison:mean_a_offset_5})
         and
         (\protect\subref{fig:pitfalls:propagation_type_comparison:osculating_a_offset_5})
         are equivalent to
         (\protect\subref{fig:pitfalls:propagation_type_comparison:mean_a})
         and
         (\protect\subref{fig:pitfalls:propagation_type_comparison:osculating_a}),
         but show the errors when the propagation is performed over
         five epochs, as opposed to one. Here the differences between
         propagating mean elements and using the full SGP4/SDP4 model
         are much starker. The propagations using the full SGP4/SDP4
         model exhibit substantially larger noise in their errors.
         \figRef{fig:pitfalls:high_frequency_inclusion} and
         \figRef{fig:pitfalls:high_frequency_inclusion_offset_5} in
         \appendixRef{sec:appendix:additional_pitfall_1} show similar
         plots to
         (\protect\subref{fig:pitfalls:propagation_type_comparison:mean_a})
         through
         (\protect\subref{fig:pitfalls:propagation_type_comparison:osculating_a_offset_5}),
         but also include the inclination and eccentricity.  }
       \label{fig:pitfalls:propagation_type_comparison}
\end{figure}

Due to the high-frequency nature of influences that affect satellite
orbits, small differences in mean orbital elements can lead to
substantial differences in resulting osculating orbits or calculated
satellite positions. We argue that the mean elements and their
propagations contain sufficient information for detecting changes in
satellite orbits and that the addition of non-Keplerian components is
unhelpful.

In the remainder of this subsection, we use the TLE data for the
Jason-2 satellite in 2013 and 2014 as an example, with maneuver data
independently obtained from the International DORIS Service (see
\autoref{sec:manoeuvre} for details).  We show that using the full
SGP4/SDP4 model for propagation and comparison of this data is less
effective than just propagating mean orbital elements. Although this
does not demonstrate that this will always be the case, it suggests
that care should be taken before employing SGP4/SDP4 when analysing
lifetime trends in satellites.

\figRef{fig:pitfalls:propagation_type_comparison} shows the results of
performing propagations and comparisons using the different techniques
described above for the TLE data of Jason 2 from 2013 and 2014. We
begin with the typical approach for detecting anomalous updates in TLE
data. This technique propagates each TLE to the timestamp of the
subsequent epoch using the full SGP4/SDP4 model.  Non-Keplerian
components are then added to the TLE at the subsequent epoch using
SGP4/SDP4. \figRef{fig:pitfalls:propagation_type_comparison:radial}
Shows the differences in the radial direction between the resulting
propagated and non-propagated
positions. \figRef{fig:pitfalls:high_frequency_positions:in-track} and
\figRef{fig:pitfalls:high_frequency_positions:cross-track} in
\appendixRef{sec:appendix:additional_pitfall_1} show the same
calculations in the cross-track and in-track directions,
respectively. There is not a particularly close alignment between the
magnitude of the errors and the manoeuvre timestamps in any of these
three directions, especially when compared with the similar plots for
the semi-major axis
(\figRef{fig:pitfalls:propagation_type_comparison:mean_a} and
\figRef{fig:pitfalls:propagation_type_comparison:osculating_a}).

Next, we perform similar computations, but find the differences between propagated and measured
Keplerian orbital elements. This is performed both for the osculating elements, using the full SGP4/SDP4 model, as well as
for the mean elements.
\figRef{fig:pitfalls:propagation_type_comparison:mean_a} shows the results of
propagating the mean semi-major axis and finding the difference with the semi-major axis
of the TLE at the subsequent epoch (obtained from the mean motion).
This can be compared with \figRef{fig:pitfalls:propagation_type_comparison:osculating_a}, which shows the results of
the same process, but using the corresponding osculating elements. That is, the TLEs are propagated using the full
SGP4/SDP4 model and the TLEs at the subsequent epoch are converted to osculating Keplerian elements using SGP4/SDP4. Although
these plots are similar, there are distinct differences between them. For instance, in June 2013 there is a single
very large propagation error in the osculating element propagations that is not present in the mean element
propagations.

Similar plots, showing the propagation errors for both mean and osculating elements, for the inclination and eccentricity
can be found in \figRef{fig:pitfalls:high_frequency_inclusion}.

Many methods for detecting anomalous updates to TLEs perform multiple propagations for each epoch, some of which will
be over multiple epochs (that is, they will propagate further than the adjacent epoch)
\cite{decoto2015technique, lemmens2014two, li2018new, li2019maneuver}.
\figRef{fig:pitfalls:propagation_type_comparison:mean_a_offset_5} and
\figRef{fig:pitfalls:propagation_type_comparison:osculating_a_offset_5} show the same plots as
\figRef{fig:pitfalls:propagation_type_comparison:mean_a} and
\figRef{fig:pitfalls:propagation_type_comparison:osculating_a}, respectively,
however, the propagation is performed over five epochs. These show that performing
propagations and comparisons using the full SGP4/SDP4 model introduces
substantially more noise as opposed to operating solely on mean elements.
\figRef{fig:pitfalls:high_frequency_inclusion_offset_5} shows plots of the propagation errors over five epochs for
other orbital elements, with the same conclusion.

In summary, for the Jason-2 satellite in the years 2013 and 2014, if
we want to propagate the TLEs and compare them against TLEs in
subsequent epochs, then there are advantages to propagating the mean
elements and comparing against the mean orbital elements recorded in
the TLE for the subsequent epoch. Using the full SGP4/SDP4 model as
well as calculating differences in satellite positions can introduce
additional noise in the calculations and produce large deviations that
do not align well with changes in the satellite's orbit.

\subsection{Pitfall 2: Sampling Osculating Orbits at Low Rates}
\label{sec:pitfalls:osculating_at_tle_timestamps}

Here we recommend that it is preferable to avoid converting TLEs to osculating orbits
or satellite positions using SGP4/SDP4 before performing further analysis. Moreover,
if this is done then care must be taken to use a sufficiently high sampling rate.

When analysing TLE data, especially when applying a general-purpose
data analysis algorithm, it seems reasonable to convert the
mean elements in the TLE data to either osculating orbital elements or
positions. Mean elements are, after all, an approximation. One can use
SGP4/SDP4 to produce such values either at regularly sampled
intervals, or at the epoch times associated with the TLE, moreover,
the most common software packages for working with TLE data do such
transformations {\em by default}, to aid in ease of use. 

It is then necessary to choose a method by which a sample point will be
placed. One option (which is explored below) is to sample at the same
time points as the published TLE epochs --- resulting in a sampling
rate of around once per day, but irregularly placed samples, which is
inconvenient for most time-series algorithms. An alternative is to
interpolate to a set of samples at regular intervals.  For example,
Roberts and Linares~\cite{roberts2021geosynchronous} sample at a fixed
interval of one day. However, both such approaches inadequately sample
the non-Keplerian components of satellite orbits. For instance, the
variation in the semi-major axis of many satellites has a dominant
frequency at twice the orbital frequency (see Appendix A). As
low-earth-orbit satellites can have periods below 2 hours, The {\em
  Nyquist sampling rate} (that would avoid aliasing problems) is well
above once per half hour in order to avoid aliasing.

\begin{figure}[ht!]
       \centering
       \begin{subfigure}[]{\linewidth}
           \includegraphics[width=\linewidth]{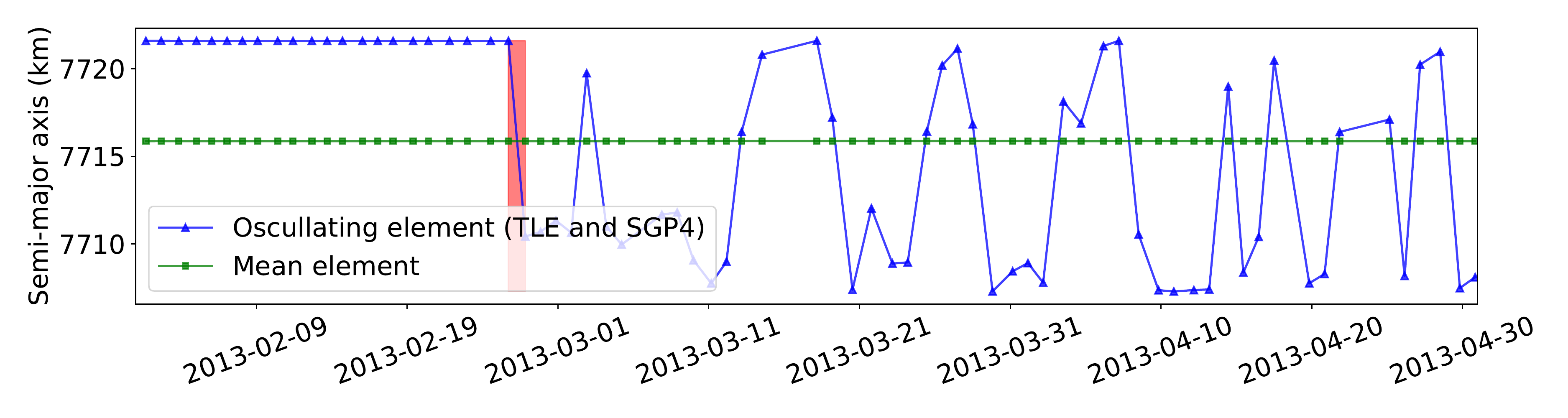}
           \caption{Semi-major axes of the mean and osculating orbits}
           \label{fig:pitfalls:jason2_oddness:mean_and_osculating}
       \end{subfigure}
       \begin{subfigure}[]{\linewidth}
           \includegraphics[width=\linewidth]{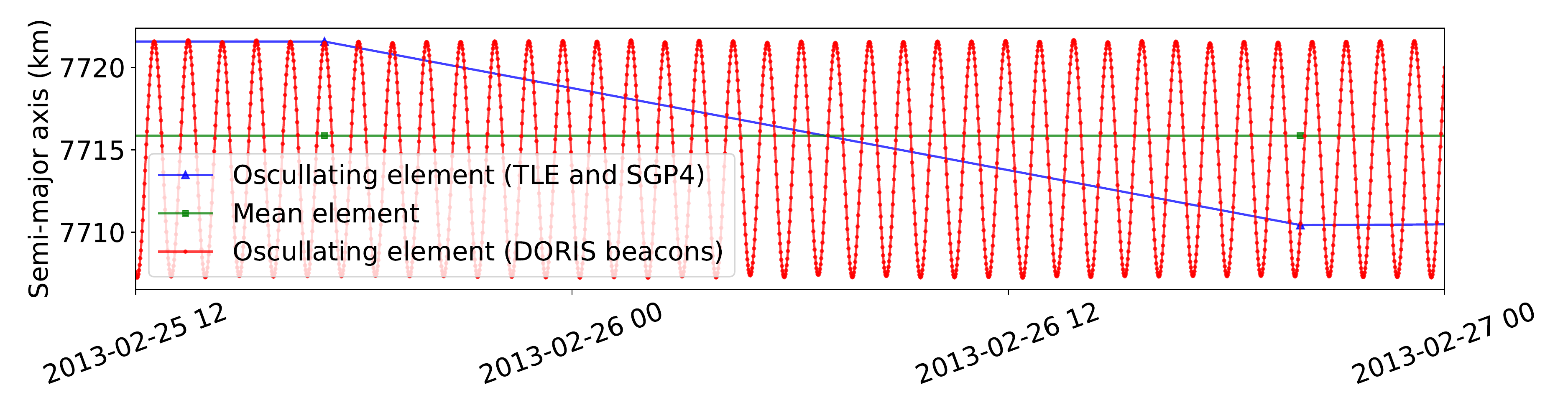}
           \caption{High-resolution view of the highlighted part of (a)}
           \label{fig:pitfalls:jason2_oddness:mean_and_osculating_and_SP3}
       \end{subfigure}
       \caption{ A vivid demonstration of how TLE data can be confusing to work with.
           (\protect\subref{fig:pitfalls:jason2_oddness:mean_and_osculating}) shows both the mean and osculating
           values for the semi-major axis of the Jason 2 satellite in early 2013. We observe that the osculating
           values change from being constant to varying substantially. This is strange behaviour which would probably
           trigger an anomaly in many anomaly detection algorithms. The causes of this strange behaviour are made
           clear in (\protect\subref{fig:pitfalls:jason2_oddness:mean_and_osculating_and_SP3}), which focuses in on
           the onset of the fluctuating period. In this figure, more accurate data derived from DORIS ground beacons
           \cite{auriol2010doris} is included in the plot. We can observe that, prior to the fluctuations, the
           timestamps associated with the TLE data aligned with the apex of the oscillations in the semi-major axis.
           This alignment was broken at the point the fluctuations begun. } \label{fig:pitfalls:jason2_oddness}
\end{figure}

The TLE data of the Jason 2 satellite contains a striking example of
the potential negative consequences of analysing osculating elements
with low sampling rates. The semi-major axis of the osculating orbit
of Jason 2 is plotted (in blue) in
\figRef{fig:pitfalls:jason2_oddness:mean_and_osculating}, along with
the semi-major axis of the mean orbit (in green). The figure includes
all epochs between the 1\textsuperscript{st} of February and the
1\textsuperscript{st} of May 2013. Between the 1\textsuperscript{st}
and 25\textsuperscript{th} of February, the semi-major axis of the
osculating orbit appears to remain almost constant. After the
25\textsuperscript{th} of February it begins fluctuating
substantially.

Observing only the semi-major axis of the osculating orbit, many
reasonable observers would be tempted to conclude that some feature of
the satellite's orbit had changed on or around the
25\textsuperscript{th} of February 2013. However, we can also see that
the semi-major axis of the mean orbit undergoes no substantial change
and there was no published change in the orbit or mission of Jason 2
during this period.

\begin{figure}[ht!]
       \centering
       \begin{subfigure}[]{\linewidth}
           \includegraphics[width=0.8\linewidth]{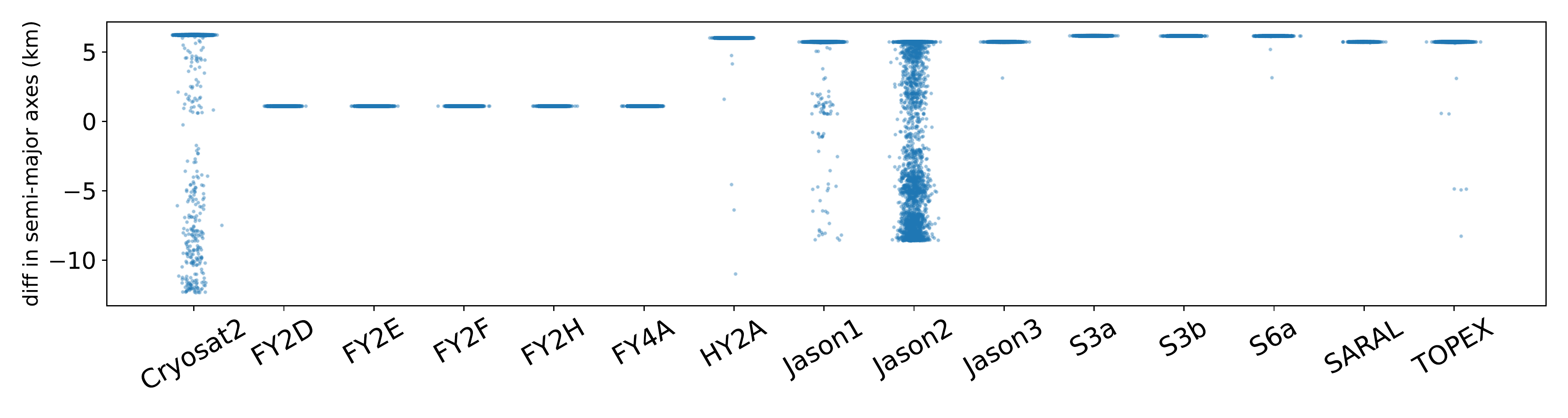}
           \caption{Semi-major axis}
           \label{fig:pitfalls:diffs_mean_and_osculating:sma}
       \end{subfigure}
       \begin{subfigure}[]{\linewidth}
           \includegraphics[width=0.8\linewidth]{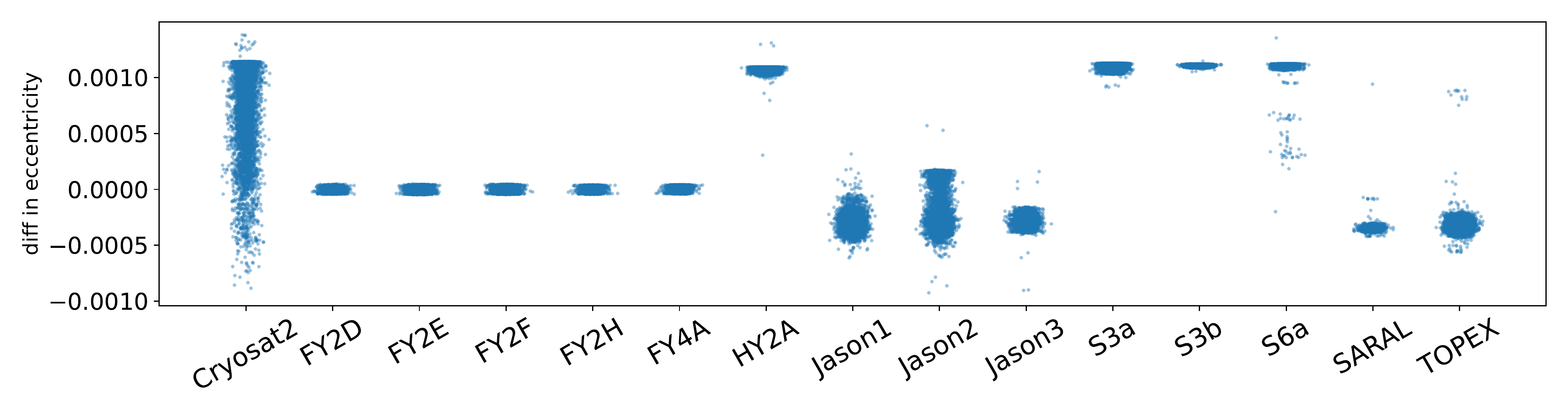}
           \caption{Eccentricity}
           \label{fig:pitfalls:diffs_mean_and_osculating:ecc}
       \end{subfigure}
       \begin{subfigure}[]{\linewidth}
           \includegraphics[width=0.8\linewidth]{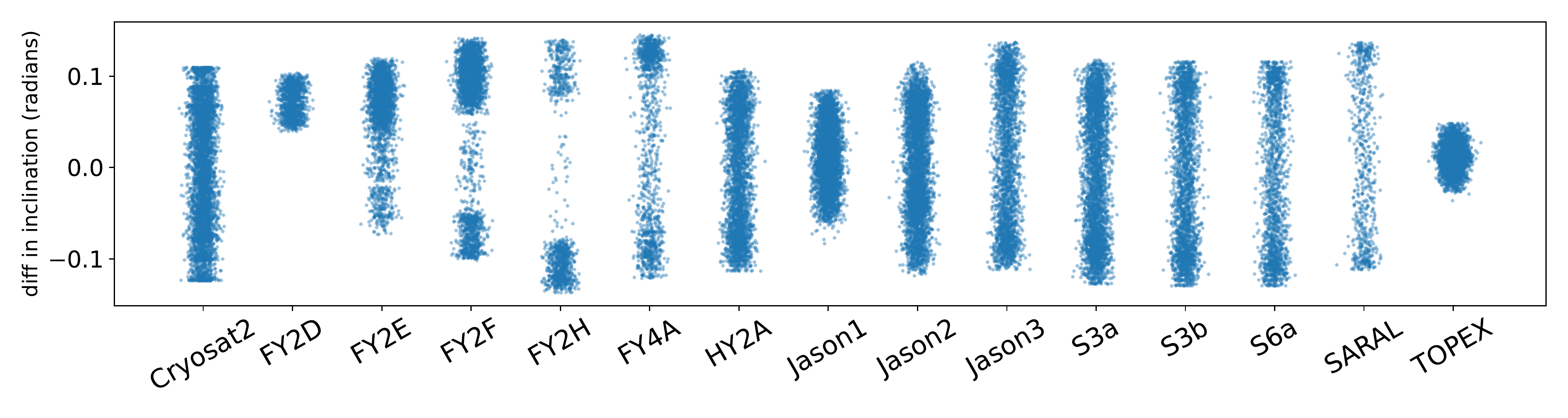}
           \caption{Inclination}
           \label{fig:pitfalls:diffs_mean_and_osculating:inc}
       \end{subfigure}
       \caption{Rainfall plots visualising the distributions of the differences between orbital elements of the mean
       and the osculating orbits of the satellites included in the
       presented dataset.
For the semi-major axis (\protect\subref{fig:pitfalls:diffs_mean_and_osculating:sma}), the majority of
       satellites have highly
       concentrated distributions. This is due to the epoch times always coinciding with the same phase in the
       high-frequency parts of the non-Keplerian components of the orbit. Other satellites, notably Cryosat 2, Jason 1
       and Jason 2 have much
       broader distributions in this difference. Inspection of the semi-major axis of the osculating orbit
           (as show in \figRef{fig:pitfalls:jason2_oddness:mean_and_osculating}), shows that, in these cases, the phase
       within the non-Keplerian components of the orbit at which the epoch times are published varies during the period
       of the published data. There is a similar effect for the eccentricity
       (\protect\subref{fig:pitfalls:diffs_mean_and_osculating:ecc}), where most satellites have a highly
           concentrated distribution, whereas Cryosat and the Jasons have broader concentrations. Conversely, all
           satellites have similarly broad distributions for the inclination
           (\protect\subref{fig:pitfalls:diffs_mean_and_osculating:inc}), indicating that the published epoch times
           are not sampling this component synchronously (that is, always at the same point within an oscillation).
       }
       \label{fig:pitfalls:diffs_mean_and_osculating}
\end{figure}

The artefacts in the orbital data can be understood by inspecting the
highly-accurate and finely-sampled recordings of this satellite's
semi-major axis.
\figRef{fig:pitfalls:jason2_oddness:mean_and_osculating_and_SP3}
contains the semi-major axis of the osculating orbit of Jason 2 as
measured by ground radio beacons from the DORIS network
\cite{auriol2010doris}. Due to the high-frequency variation of this
data, we have limited the $x$-axis to only cover the period from the
25\textsuperscript{th} to the 27\textsuperscript{th} of February
(where the interesting change occurred). We see that the semi-major
axis of the osculating orbit derived from the TLE data closely matches
the semi-major axis of the osculating orbit derived from the DORIS
beacons, both before and after the change in variation. Only two TLE
data points can be shown in this plot, but this close correspondance
continues to both the left and the right of the plotted region. What
does change, however, is that the epoch dates published by NORAD align
with the apex of the oscillations before the 25\textsuperscript{th} of
February but this alignment is broken after this date. In essence, the
manner in which the high-frequency parts of the non-Keplerian
components are being aliased changes.

The conclusion of this analysis of Jason 2 is that, if we sample the osculating orbit at the published epoch times,
we can get a very skewed view of the actual changes in the orbit of the satellite. In this example,
although we observe a large change in the observed data after the 25\textsuperscript{th} of February 2013, this does
not reflect any corresponding large change in the orbit of the satellite. It is merely the result of a relatively small
change made by NORAD in times at which is observed the data.

\figRef{fig:pitfalls:diffs_mean_and_osculating} shows the degree to which this effect is present in the historic data
for the various satellites included in the benchmark dataset presented in \secRef{sec:dataset_description}. Rainfall
plots show the distribution of the differences between the orbital elements associated with the osculating and mean
orbit at each published epoch. This is done for the semi-major axis
(\figRef{fig:pitfalls:diffs_mean_and_osculating:sma}), eccentricity
(\figRef{fig:pitfalls:diffs_mean_and_osculating:ecc}) and inclination
(\figRef{fig:pitfalls:diffs_mean_and_osculating:inc}). Most satellites in this dataset have relatively narrow
distributions of these differences for the semi-major axis and eccentricity. This is indicative
of the published epoch times sampling these components synchronously. That is, they are always
sampled at (close to) the same point in their phase. By contrast, CryoSat-2 and the Jason satellites have much
broader distributions, which indicates changes in the phase at which
this sampling takes place. However, the important point is that we
have no control over exact sample times, and so these same effects
could be present in any satellite's data at some time period.

Aliasing is a well-understood phenomena in signal processing, but is
perhaps so well-known that most signal processing data are
automatically passed through band-pass filters before sampling. It is
therefore commonly assumed that a dataset will have had aliasing
effects removed before proceeding to the stage of, for instance,
change detection. It is therefore a key pitfall that certain aspects
of TLE data have not undergone this pre-processing.

\subsection{Pitfall 3: Comparing Full SGP4 Propagations to Mean Elements}
\label{sec:pitfalls:mean_to_propagations}

In this subsection we recommend that, if TLEs are propagated using the full SGP4/SDP4 model, then care must be taken
to ensure that TLEs in subsequent epochs are also transformed with this model before comparisons are made.

There is a further trap relating to propagating TLEs. It is similar in nature to the first pitfall discussed in
\secRef{sec:pitfalls:full_sgp4}, however, it can lead to more serious consequences. This occurs
when the TLEs are propagated using the full SGP4/SDP4 model, but the resulting osculating orbital elements are compared to
the mean TLE elements in the subsequent epoch. That is, the non-Keplerian components of the orbit are not added to the
TLE at the subsequent epoch before the comparison is made. The description of TLE data and the SGP4/SDP4
model provided in \secRef{sec:TLE_description} reveals why this is problematic --- it results in a
comparison between fundamentally different types of orbital elements.

\begin{figure}[ht!]
       \centering
       \begin{subfigure}[]{\linewidth}
         \includegraphics[width=\linewidth]{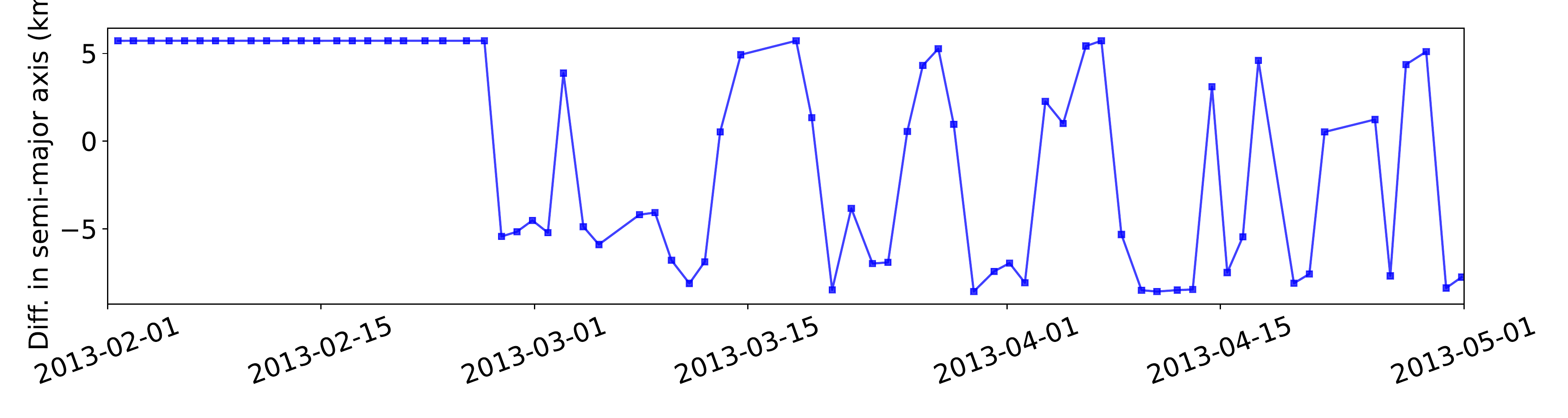}
           \caption{Differences between the semi-major axis propagated from the previous epoch and the mean element
           at the current epoch.}
           \label{fig:pitfalls:jason2_comparing_propagations_with_mean:mean}
       \end{subfigure}
       \begin{subfigure}[]{\linewidth}
           \includegraphics[width=\linewidth]{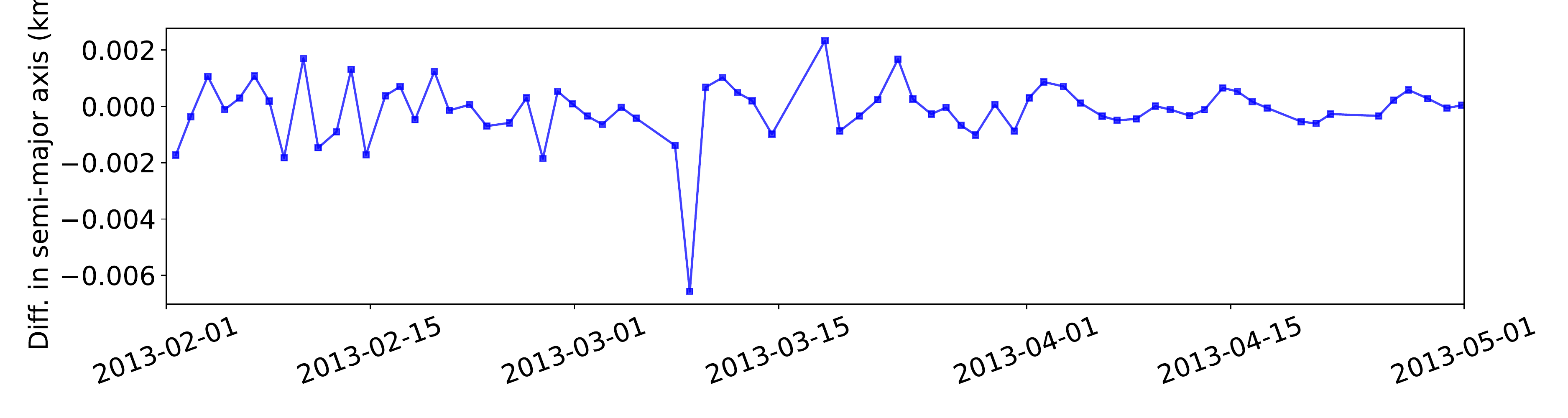}
           \caption{Differences between the semi-major axis propagated from the previous epoch and the osculating
           element at the current epoch.}
           \label{fig:pitfalls:jason2_comparing_propagations_with_mean:osculating}
       \end{subfigure}
       \caption{A demonstration of why it is necessary to be careful what propagated TLE elements are compared
       against. Both plots show the differences between the semi-major axis propagated from the previous epoch to the
       current epoch and that recorded at the current epoch for the Jason 2 satellite. Note that the time period
       covered is the same as in \figRef{fig:pitfalls:jason2_oddness:mean_and_osculating}.
       Subfigure~(\protect\subref{fig:pitfalls:jason2_comparing_propagations_with_mean:mean}) shows the difference with the
       mean element and (\protect\subref{fig:pitfalls:jason2_comparing_propagations_with_mean:osculating}) shows
       the difference with the osculating element at the current epoch. Note the difference in the scales of the $y$
           axes. Two points
       should stand out. The first is that the differences between the propagated element and the osculating element are
       substantially lower than the differences between the propagated element and the mean element. The second
       is that, when comparing  with the mean element, the distribution of the differences goes through a dramatic
       change in early March of 2013. No such change is observed when comparing with the osculating element.}
       \label{fig:pitfalls:jason2_comparing_propagations_with_mean}
\end{figure}

The consequence of this is that changes in the differences between propagated and subsequent elements might
be dominated by changes in the phase at which the high-frequency parts of the non-Keplerian components are sampled, as
opposed to changes in the orbit itself. An example of this is shown in \figRef{fig:pitfalls:jason2_comparing_propagations_with_mean},
where \figRef{fig:pitfalls:jason2_comparing_propagations_with_mean:mean} shows the result of propagating the TLEs of Jason
2 using the full SGP4/SDP4 model and comparing the resulting osculating semi-major axis with the mean semi-major axis
in the subsequent epoch. Note that
\figRef{fig:pitfalls:jason2_comparing_propagations_with_mean:mean} is plotting the same satellite over the same time
period as
\figRef{fig:pitfalls:jason2_oddness:mean_and_osculating}. We observe a similar effect to the
second pitfall described in \secRef{sec:pitfalls:osculating_at_tle_timestamps}: the residuals between the
propagations and the TLEs undergo a substantial increase after the 25\textsuperscript{th} of February 2013. As in
that instance, it would be tempting to conclude that there was a change in the satellite's orbit. However, as
discussed in \secRef{sec:pitfalls:osculating_at_tle_timestamps}, there are no published changes in the orbit or
mission of Jason 2 around this time. Moreover, close inspection of accurate orbital data
(see \figRef{fig:pitfalls:jason2_oddness:mean_and_osculating_and_SP3}) reveals that what has changed is that the
published epoch times no longer coincide with the peaks in the oscillations of the semi-major axis. This leads to a
change in the variance of the osculating semi-major axis sampled at these points, which subsequently changes the
variance of the residuals. \figRef{fig:pitfalls:jason2_comparing_propagations_with_mean:osculating} plots the
differences between the propagated semi-major axis (using the full SGP4/SDP4 model) and the osculating semi-major axis
obtained from the subsequent TLE. Although this approach will suffer from the issues highlighted in
\secRef{sec:pitfalls:full_sgp4}, during analysed period it does not experience the same dramatic
changes as comparing osculating and mean elements.
 
\FloatBarrier

\section{Dataset Description}
\label{sec:dataset_description}

There now exists a substantial body of work describing methods for
analysing trends in a satellite's orbit across its lifetime using TLE
data. See \secRef{sec:previous_work} for a review of this literature.
It is common, however, for each contribution to evaluate its methods
on different satellites; and little attempt is therefore made to make
performance comparisons with previously approaches. Evaluation is also
usually performed on a relatively small set of satellites (2-5), and
ground-truth data, against which performance is measured is rarely
obtained from an independent source.

We present here a curated, benchmark dataset for satellite orbit
surveillance using TLE data. The goal is to facilitate future studies
by providing and easy-to-use dataset, that is larger than any existing
set, and provides accurate ground truth for better performance
evaluations and comparisons.  The dataset consists of 15 diverse
satellites and associated manoeuvre timestamps.
It can be found at
\texttt{\url{github.com/dpshorten/TLE_observation_benchmark_dataset}}.

The core of the curated dataset consists of the TLE data and
associated independently obtained ground-truth manoeuvre
timestamps. More accurate DORIS data detailing precise satellite positions
(as available) is also included. This additional data is only intended
to be used for the purpose of determining the accuracy of the TLE data
and the associated SGP4/SDP4 propagator \cite{celesTrak}.  A list of
all satellites included in the dataset is given in
\tableRef{tab:dataset_description:list_summary}, along with summary
statistics of the data and features of the satellites. These
satellites were chosen based on availability of the manoeuvre
timestamps, as well as with the goal of providing a diversity of types
of satellite, including both geostationary and low-earth-orbit
satellites.

\begin{table}[!ht]
    \setlength\tabcolsep{2mm}
    \centering

    \begin{tabular}{p{2.3cm}|S[table-format=5.0]|S[table-format=4.0]|S[table-format=3.0]|
    p{2cm}|p{2cm}|c|S[table-format=5.1]}
        & \textbf{SATCAT} & \textbf{TLE} &
      \multicolumn{3}{c|}{\textbf{Observed Manoeuvres}} &  & \textbf{Altitude} \\ \cline{4-6}
        \textbf{Name} & \textbf{Number} & \textbf{Epochs}  & \textbf{Num} & \textbf{First} & \textbf{Last} &
        \textbf{DORIS} & \textbf{(km)}\\\hline
        Fengyun-2D     & 29640    & 1178   & 22     & 2011-02-01   & 2015-04-10   &               & 35786 \\
        Fengyun-2E     & 33463    & 2367   & 48     & 2011-03-17   & 2018-10-15   &               & 35786 \\
        Fengyun-2F     & 38049    & 2977   & 68     & 2012-09-11   & 2022-01-05   &               & 35786 \\
        Fengyun-2H     & 43491    & 1043   & 12     & 2019-01-18   & 2022-01-18   &               & 35786 \\
        Fengyun-4A     & 41882    & 1296   & 49     & 2018-05-22   & 2022-02-21   &               & 35786 \\
        Sentinel-3A    & 41335    & 2226   & 64     & 2016-02-23   & 2022-10-07   & \checkmark    & 814.5 \\
        Sentinel-3B    & 43437    & 1416   & 56     & 2018-05-01   & 2022-10-07   & \checkmark    & 814.5 \\
        Sentinel-6A    & 46984    & 566    & 18     & 2020-11-24   & 2022-10-14   & \checkmark    & 1336 \\
        Jason-1        & 26997    & 4000   & 119    & 2001-12-12   & 2013-06-14   & \checkmark    & 1324 \\
        Jason-2        & 33105    & 3946   & 111    & 2008-06-24   & 2019-10-05   & \checkmark    & 1309.5 \\
        Jason-3        & 41240    & 2315   & 43     & 2016-01-20   & 2022-10-11   & \checkmark    & 1336 \\
        SARAL          & 39086    & 3165   & 62     & 2013-02-28   & 2022-09-22   & \checkmark    & 800 \\
        CryoSat-2      & 36508    & 4243   & 168    & 2010-04-16   & 2022-10-06   & \checkmark    & 717 \\
        Haiyang-2A     & 37781    & 3025   & 58     & 2011-09-29   & 2020-06-10   & \checkmark    & 975 \\
        TOPEX          & 22076    & 4156   & 43     & 1992-08-18   & 2004-11-18   &               & 1336 \\
    \end{tabular}
    \caption{List of the satellites in the curated dataset,
         their SATCAT Number \cite{CelestrakFAQ}, and summary statistics. The first and
        last manoeuvres only refer to those included in the dataset.
        The ``DORIS'' column signifies whether the dataset includes the accurate position
        data derived from the DORIS ground beacons \cite{auriol2010doris} for each satellite.
        The altitude values were obtained from the World Meteorological Association OSCAR
        database \protect\footnotemark.}
    \label{tab:dataset_description:list_summary}
\end{table}

\begin{figure}[ht!]
       \centering
       \includegraphics[width=\linewidth]{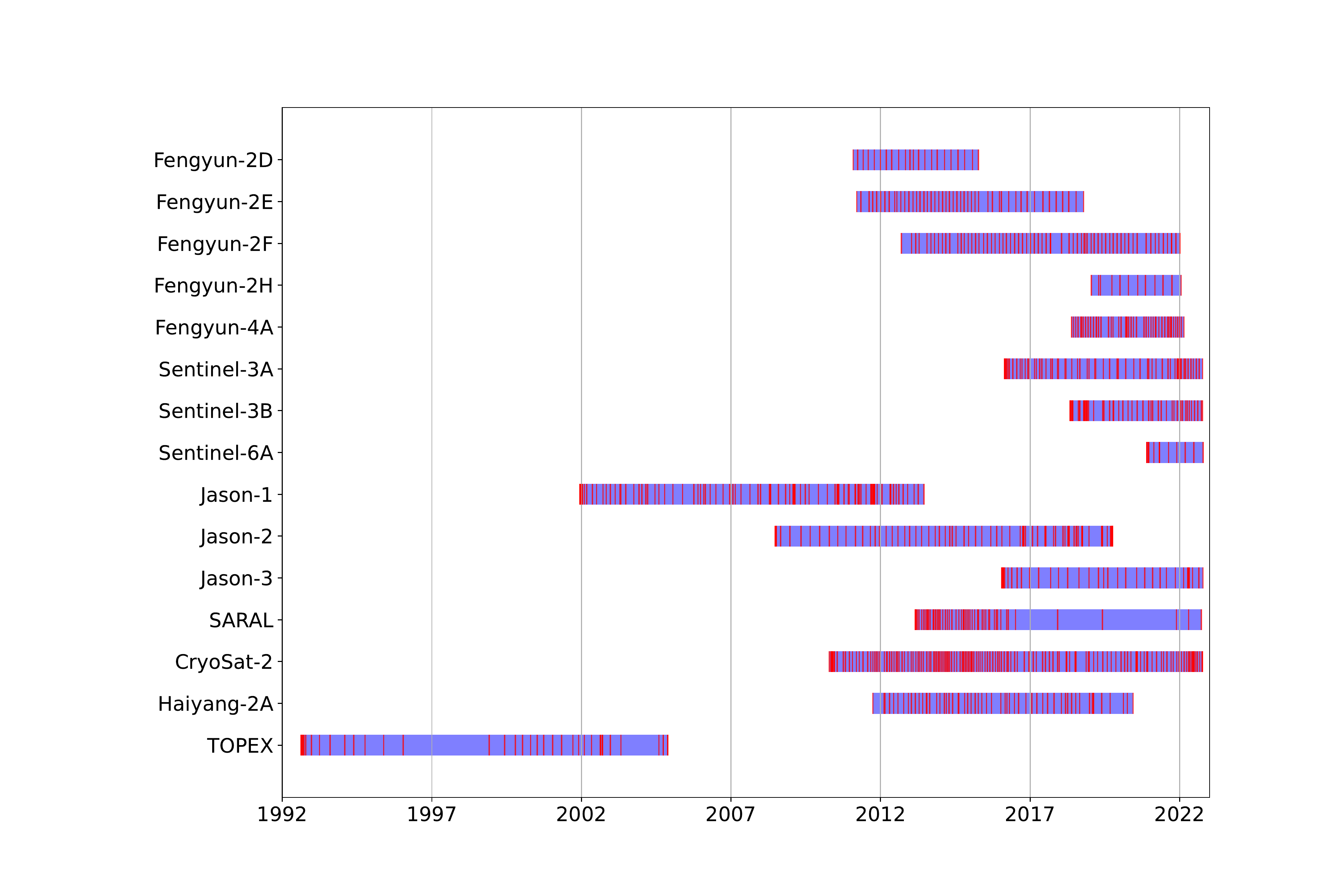}
       \caption{Time periods of satellite observation data sets. Red lines show the
       timestamps of each manoeuvre. For each satellite, the left point corresponds closely to its launch. The right
       point either corresponds with its end of life, or when the data was accessed by the authors.}
       \label{fig:dataset_description:satellite_gantt}
\end{figure}

\begin{figure}[ht!]
       \centering
       \begin{subfigure}[]{\linewidth}
           \centering
           \includegraphics[width=0.6\linewidth]{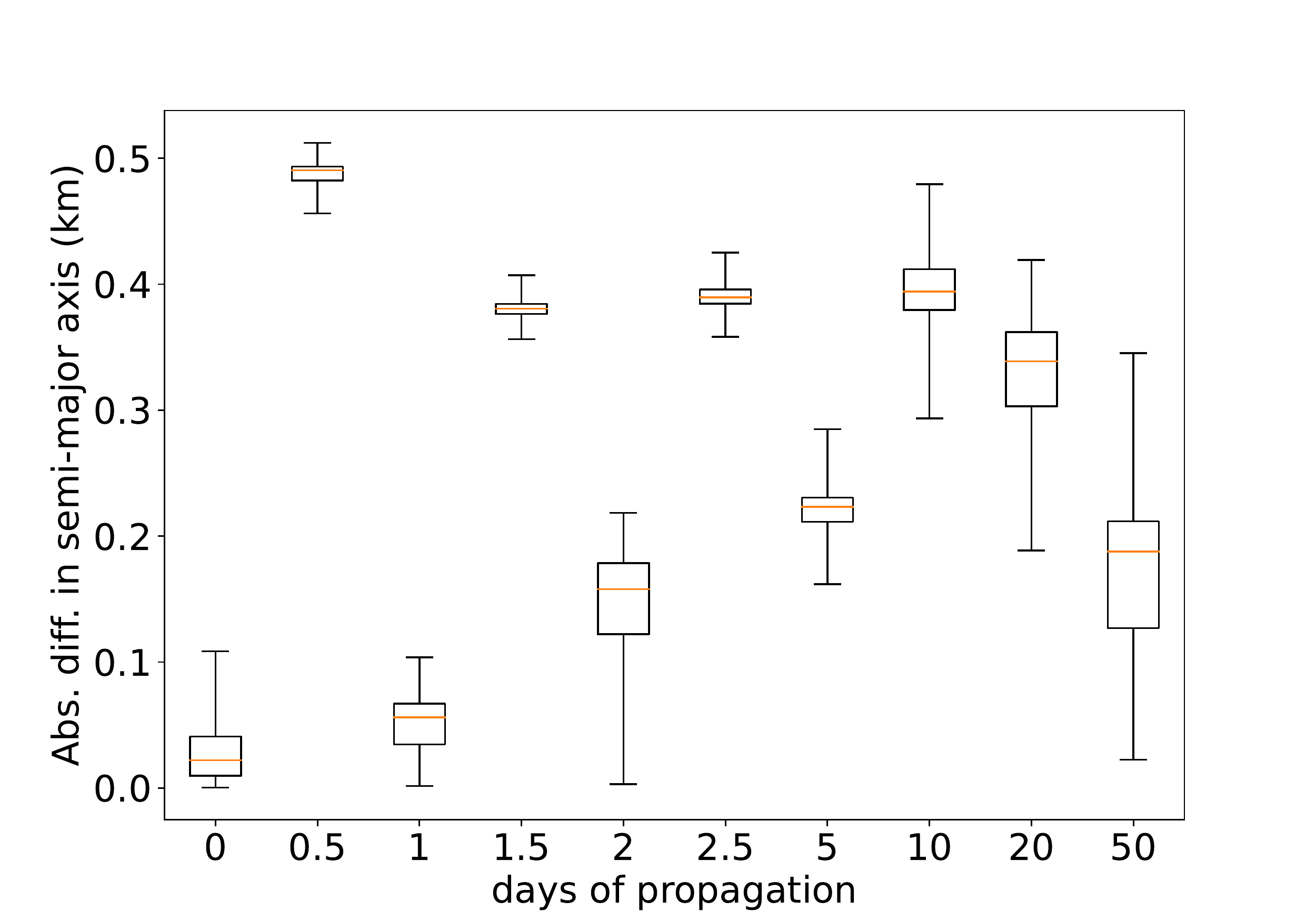}
           \caption{Sentinel-3A}
           \label{fig:dataset_description:propagation_errors:S3A}
       \end{subfigure}
       \begin{subfigure}[]{\linewidth}
           \centering
           \includegraphics[width=0.6\linewidth]{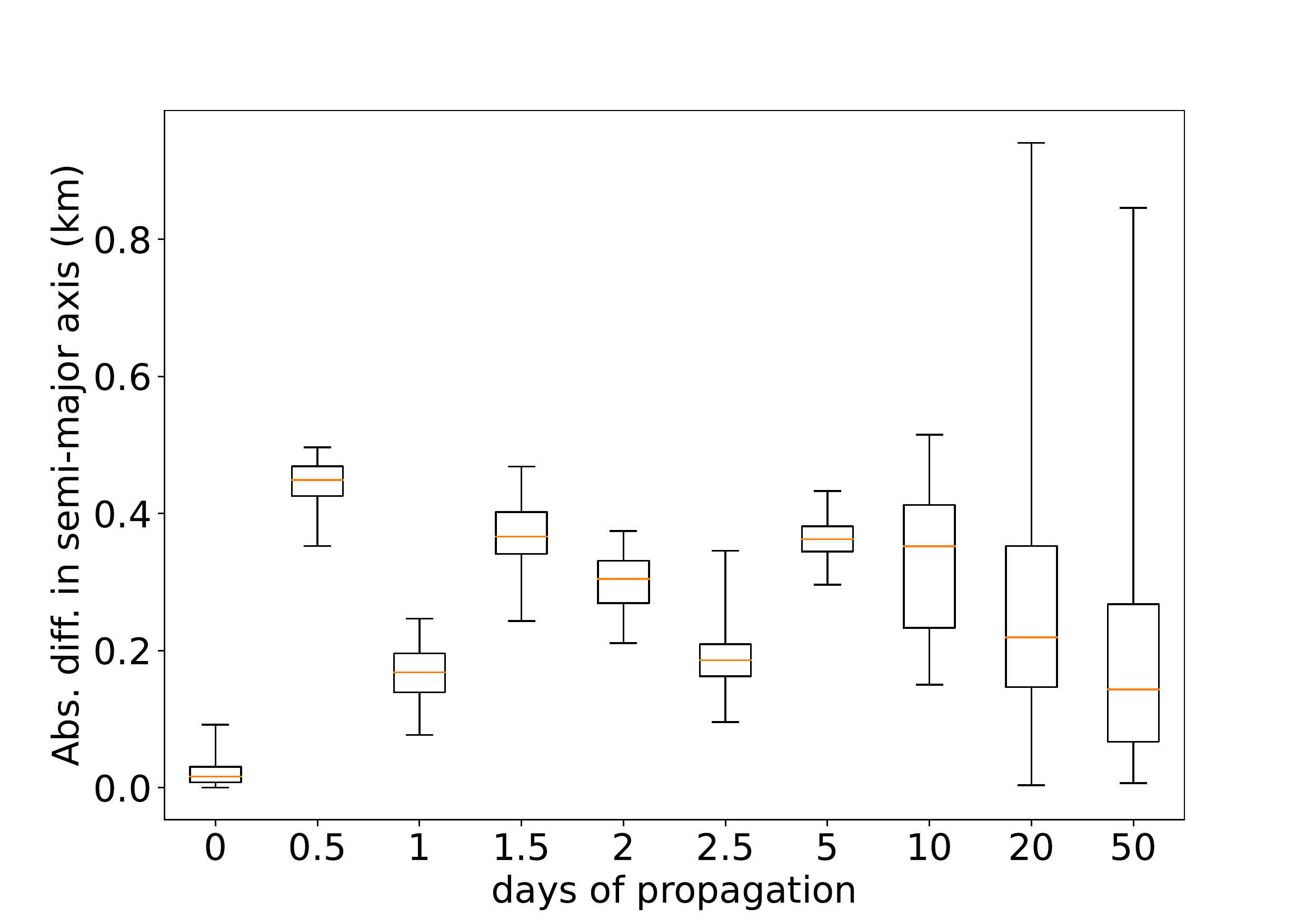}
           \caption{Haiyang-2A}
           \label{fig:dataset_description:HY2A}
       \end{subfigure}
       \caption{Box plots showing the distribution of propagation errors using SGP4/SDP4. Randomly chosen TLEs are
       propagated
       for the number of days shown on the $x$ axis. The propagated values are then compared with accurate data
       derived from DORIS ground beacons \cite{auriol2010doris}. The large fluctuations in the mean of the
       propagation errors for small numbers of days of propagation is due to the known periodicity in the propagation
       errors of SGP4/SDP4 \cite{aida2013accuracy}. This periodicity makes it difficult to suggest a cutoff point
       beyond which the accuracy of the propagator is considered insufficient.}
       \label{fig:dataset_description:propagation_errors}
\end{figure}

\subsection{Manoeuvre Timestamps}
\label{sec:manoeuvre}

\footnotetext{\texttt{space.oscar.wmo.int} accessed: 26 October 2022}

The manoeuvre timestamps are stored in YAML files \cite{YAML}. These files also contain the satellite name and SATCAT catalogue
number \cite{CelestrakFAQ} as metadata. For all satellites other than the Fengyung satellites, these were obtained
from the International DORIS Service.
Specifically, they were downloaded from \texttt{ftp.ids-doris.org/pub/ids/satellites/} on the 24th of October 2022.
Documentation on file formats can be obtained at that location.
Note that, although there is no ground-beacon data available for the TOPEX
satellite from the International DORIS Service, its manoeuvres were obtained from this source.
Manoeuvre timestamps were extracted from the downloaded files and packaged into the YAML files in this dataset.
Orbit control information for Fengyung satellites is provided by China's National Satellite Meteorological Center on
its official website \texttt{www.nsmc.org.cn/nsmc/en/news/index.html}.
These manoeuvre events have been collected and shared by colleagues from Purple Mountain Observatory.
\tableRef{tab:dataset_description:list_summary} shows the dates of the first and last manoeuvres for
each satellite in the dataset, along with the number of manoeuvres. The period spanned by the manoeuvre data
for each satellite is also visualised in \figRef{fig:dataset_description:satellite_gantt}.

\subsection{TLE Data}

The TLE data is stored in plain-text files in accordance with the specification provided by NORAD \cite{celesTrak}.
This data was downloaded from \texttt{space-track.org} (a service provided by NORAD) on 26 October 2022. The files
contain all TLE data from 1 week before the first manoeuvre in the associated manoeuvre file until 1 week after the
last manoeuvre. \tableRef{tab:dataset_description:list_summary} shows the number of TLE epochs present in the data
for each satellite. This TLE data is easily obtained. We incorporate it as part of this benchmark dataset in order
to provide a standardised dataset (particularly with respect to start and end dates) on which to test methodologies for
space situational awareness.

\subsection{Accurate Satellite Position Data}

The benchmark dataset also contains highly accurate (relative to TLE)
satellite positions for a subset of satellites. This data is
obtained from the DORIS network of ground beacons
\cite{auriol2010doris}, where it was available. These beacons emit radio signals which are
received by the satellites. Satellite positions can then be determined from the Doppler shift of the received signals.
The derived satellite positions are highly accurate (on the order of centimeters) and published at 1 minute intervals.

This data was obtained from \texttt{doris.ign.fr/pub/doris/products/orbits/} on the 28th of September
2022. The availability of this data
for each satellite is shown in \tableRef{tab:dataset_description:list_summary}.

The original data is stored in files with the SP3-c format \cite{Hilla2010Extended}. As this file format requires
specific software to manipulate \footnote{eg: pypi.org/project/sp3/0.0.2/}, we have made it available in a more
accessible format. Moreover, the positions in this data are recorded in the
Geocentric Celestial Reference Frame \cite{Petit2010IERS}. They were converted into Keplerian elements
\cite{Meeus1991astronomical}, before being written to CSV files. These files contain the timestamp of each record,
along with the osculating Keplerian elements.

\subsection{Evaluation of SGP4/SDP4 Accuracy in this Dataset}

As the SGP4/SDP4 propagation model \cite{vallado2006revisiting} is extensively used for anomaly detection applied to
TLE data, it is valuable to establish its accuracy for use in conjuction with this dataset. In particular, we are
interested in how its accuracy deteriorates over propagation length. To gain an understanding of the
accuracy of this propagator, we investigated the propagation errors on two satellites for which the high-precision
DORIS beacon data is available \cite{auriol2010doris}. For each of these satellites, 100 TLE epochs (which fell
within the range spanned by the DORIS beacon data) were chosen randomly. At each randomly chosen epoch, the timestamp
in the DORIS data that is closest to the given epoch is found. This timestamp is taken to be the starting point of
the propagation. As the TLE data is sampled around once per day and the DORIS data is sampled every minute, the
chosen timestamp in the DORIS data will align closely with the TLE epoch. SGP4 is then used to convert the mean
orbital elements in the TLE data into osculating elements, at the chosen timestamp. These osculating elements are
then compared with osculating elements obtained by converting the DORIS positions into osculating Keplerian elements.
SGP4/SDP4 is also used to propagate the TLE data to timepoints at set intervals in the future from the originally
found DORIS timestamp. Comparisons with the DORIS data are also made at these time points.

Plots of the distributions of these propagation errors are shown in
\figRef{fig:dataset_description:propagation_errors}. One item that
stands out in these plots is that, although the mean propagation error
increases with the length of the propagation, this increase is not
monotonic. Indeed, the mean error undergoes some sharp drops as we
increase the propagation length.  This is due to the known periodicity
in SGP4/SDP4 errors along propagation length \cite{aida2013accuracy,
  easthope2015examination, bai2012periodicity}. The conclusion is that
some additional care is needed in multiple-time-step propagation, but
that it can potentially be very useful, even up to fairly long time
steps.
 
\FloatBarrier

\section{Discussion}

As described in \secRef{sec:TLE_description}, TLE data is intended to be used in conjunction with the full
SGP4/SDP4 propagation model \cite{CelestrakFAQ, vallado2006revisiting, vallado2012two}. This makes sense for
the majority of the cases where one might be using TLE data. For instance, if we want to know where in the night sky
to look for a satellite, we need an accurate position for the satellite at the current instant. This
position cannot be arrived at from the mean elements in the TLE data used in isolation.

However, if we are interested in the longer-term trends in a satellite's orbit, such as whether the shape of its
orbit on a given day is different from this shape on the previous day, then the SGP4/SDP4 model is not only of far
less value, but it has the potential to be the cause of misleading results. We argue that, if one is
interested in analysing long-term changes in satellite orbits, then the full SGP4/SDP4 model should be ignored. It is
intended for a different task: predicting relatively precise satellite positions at a given instant in time. Rather,
one should focus on the mean orbital elements as recorded in the raw TLE values. Propagation of these values,
according to some model of orbital mechanics, can
be a useful step in an analysis pipeline, however, we suggest that this should be done using some method for
propagating only mean orbital elements. Indeed, SGP4/SDP4 first propagates the mean orbital elements (see page 12 of
\cite{hoots1980spacetrack}), before adding in non-Keplerian terms. Using just this component of the model would be
appropriate.

\secRef{sec:pitfalls:full_sgp4} examined specific potential negative consequences of utilising
SGP4/SDP4 for the long-term analysis of satellite orbits. It focussed on the use of SGP4/SDP4 for
propagating TLE values to the epoch time of another satellite in order to compare values. It was shown how using
the full SGP4/SDP4 model in this context could magnify the deltas in unpredictable ways (as compared to just
propagating the mean elements). \secRef{sec:pitfalls:osculating_at_tle_timestamps} looked at the use of SGP4/SDP4
in order to incorporate non-Keplerian components of the orbit before applying subsequent analysis. It was
demonstrated how, if this approach was followed, one would need to ensure that the resulting time
series were sampled at a rate much higher than the original TLE data to avoid aliasing. Moreover, a specific
case where subtle changes in how NORAD chose epoch times had a drastic impact on the osculating elements at TLE
epochs was highlighted. Finally, \secRef{sec:pitfalls:mean_to_propagations} looked at the potentially severe
scenario where TLEs propagated using the full SGP4/SDP4 model are compared against raw TLE values in subsequent
epochs. Not only is this not an apples-to-apples comparison, but it can also cause the introduction of errors due to
aliasing effects.

While our analysis has not shown that the addition of non-Keplerian components to TLE data using SGP4/SDP4 will lead
to negative consequences in all cases when analysing long-term satellite trends, we do not believe that it is
necessary to do so. It is possible that there might be positive effects to incorporating these terms for some
satellite. However, we argue that, in any data analysis task, the default approach should be to analyse the raw
records. Transformation of these raw records according to some model of dynamics should only be done if this
transformation can be demonstrated to provide value. This paper calls into question the value that is provided by
transforming TLE data using SGP4/SDP4 when analysing satellite lifetimes. Practitioners should be hesitant to use it
for this task until someone can demonstrate its effectiveness.

As summarised in \secRef{sec:previous_work}, there is
a large body of existing work on methods for observing changes in satellite orbits over long periods of time from TLE
data. However, as different studies choose different satellites on which to evaluate their methods, it is very
difficult to compare these approaches. Moreover, it is common for approaches to only be validated on one or two
satellites, thus limiting the conclusions that can be made about their effectiveness. To remedy this situation, we
have provided an open, curated benchmark dataset which contains the TLE data and associated ground-truth manoeuvre
timestamps for a set of 15 satellites. This dataset is described in \secRef{sec:dataset_description}.

\section{Conclusion}

In this paper, we studied the task of analysing trends in satellites' orbits using TLE data. We demonstrated how using
the SGP4/SDP4 orbital propagator for this task, as is regularly done in the literature, can cause significant
problems. We have argued that a more fruitful approach would be to propagate the mean orbital elements contained in
the TLE records. We have also presented and described an open curated dataset of TLE records and ground-truth
manoeuvre timestamps. It is hoped that this new dataset will facilitate future research in this area.

One main avenue for future research is to expand the presented dataset to include more satellites. The authors also
hope to investigate the use of mean element propagation for the purpose of detecting anomalies in TLE data.
 
\section{Acknowledgement}

We thank Ms Can Zhu and Dr Tinglei Zhu from Key Laboratory for Space Object and Debris Observation,
Purple Mountain Observatory, Chinese Academy of Sciences for kindly sharing Fengyun manoeuvring data. We thank
Will Heyne of BAE Systems for helpful insights throughout the course of this research. Funding for this work was
provided by the SmartSat CRC, project P2.11 . 
\begin{appendices}

\section{Frequency-space Analysis of High-Resolution Data}
\label{sec:appendix:fourier}

\begin{figure}[ht!]
       \centering
       \includegraphics[width=\linewidth]{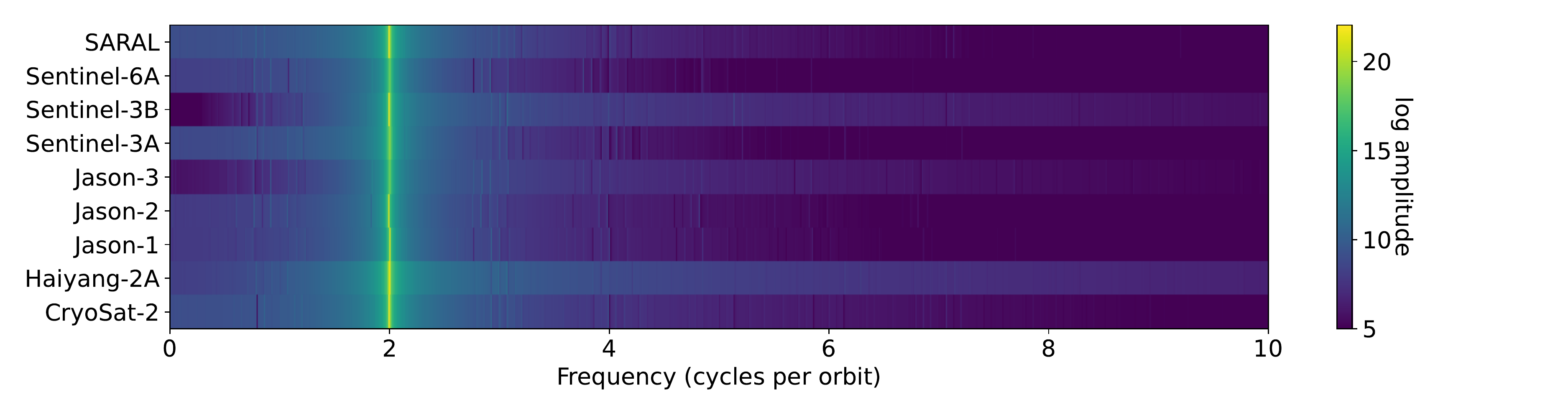}
       \caption{Fourier power spectrum of the semi-major axis of a sample of satellites from the presented dataset. The
       data used for this figure is derived from DORIS ground beacons \cite{auriol2010doris}}
       \label{fig:dataset_description:fourier_sma}
\end{figure}

\begin{figure}[ht!]
       \centering
       \includegraphics[width=\linewidth]{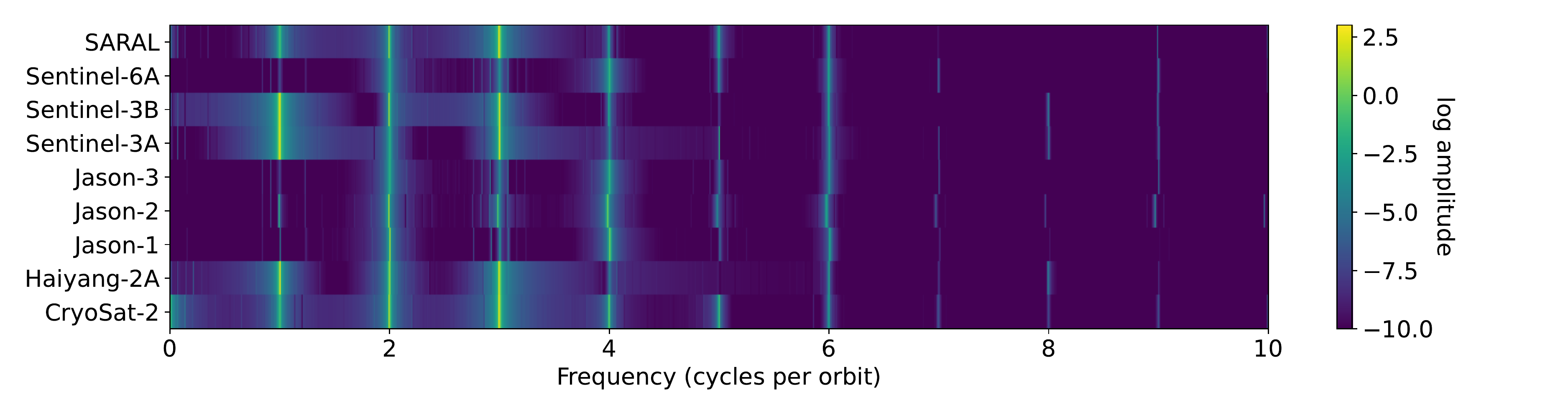}
       \caption{Fourier power spectrum of the eccentricity of a sample of satellites from the presented dataset. The
       data used for this figure is derived from DORIS ground beacons \cite{auriol2010doris}}
       \label{fig:dataset_description:fourier_eccentricity}
\end{figure}

\begin{figure}[ht!]
       \centering
       \includegraphics[width=\linewidth]{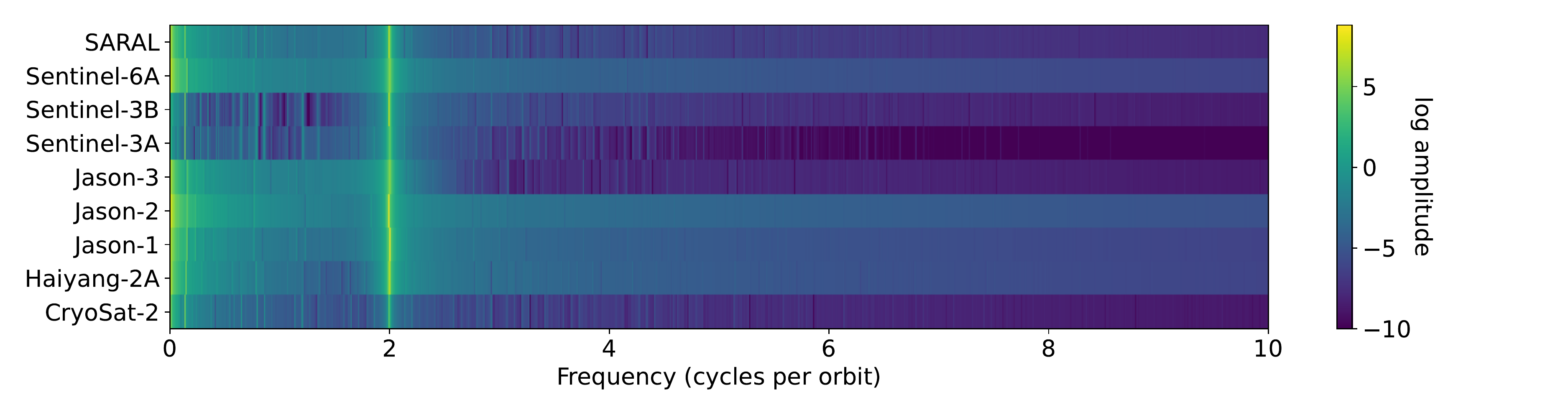}
       \caption{Fourier power spectrum of the inclination of a sample of satellites from the presented dataset. The
       data used for this figure is derived from DORIS ground beacons \cite{auriol2010doris}}
       \label{fig:dataset_description:fourier_inclination}
\end{figure}

The figures in this section show the fourier power spectrum of the osculating semi-major axis
(\figRef{fig:dataset_description:fourier_sma}), eccentricity (\figRef{fig:dataset_description:fourier_eccentricity})
and inclination (\figRef{fig:dataset_description:fourier_inclination}) for a subset of the satellites in the
presented benchmark dataset. These satellites were chosen as they are the ones for which highly accurate orbital data
from DORIS ground beacons \cite{auriol2010doris} is available. The analysis presented in these figures was
performed on that data. It is worth noting that all three of these orbital elements have dominant frequencies at twice
the orbital frequency. The orbital period of low-earth orbit satellites (which includes many satellites in the benchmark
dataset) is in the region of 2 hours \cite{sebestyen2018low}. This implies that if we sample these osculating
components at the frequency of TLE epoch updates (around once per day), we should expect these components to be
aliased. This is the root cause behind the pitfalls discussed in \secRef{sec:pitfalls:osculating_at_tle_timestamps}
and \secRef{sec:pitfalls:mean_to_propagations}.

\section{Additional Plots Concerning Pitfall One}
\label{sec:appendix:additional_pitfall_1}

\begin{figure}[ht!]
       \centering
       \begin{subfigure}[]{0.6\linewidth}
           \includegraphics[width=\linewidth]{figs/osculating_to_osculating_R_diff_offset1.pdf}
           \caption{Radial position difference}
           \label{fig:pitfalls:high_frequency_positions:radial}
       \end{subfigure}
       \begin{subfigure}[]{0.6\linewidth}
           \includegraphics[width=\linewidth]{figs/osculating_to_osculating_I_diff_offset1.pdf}
           \caption{In-track position difference}
           \label{fig:pitfalls:high_frequency_positions:in-track}
       \end{subfigure}
       \begin{subfigure}[]{0.6\linewidth}
           \includegraphics[width=\linewidth]{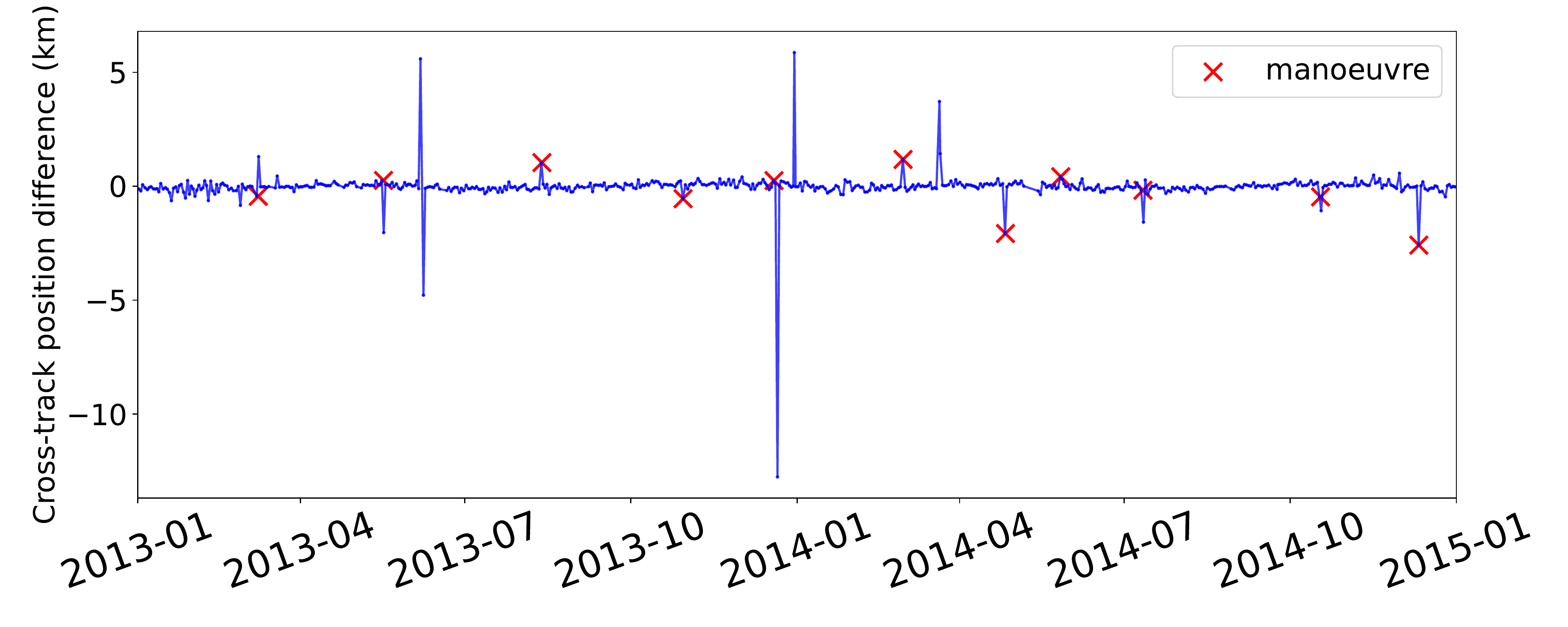}
           \caption{Cross-track position difference}
           \label{fig:pitfalls:high_frequency_positions:cross-track}
       \end{subfigure}
       \caption{
           The difference between propagated TLE positions and positions taken from the TLE in the subsequent
           epoch. The TLE from each epoch is propagated to the subsequent epoch using the SGP4/SDP4 model. SGP4/SDP4 is also
           used to incorporate non-Keplerian terms into the TLE recorded at the subsequent epoch and arrive at precise
           positions. This analysis is plotted for the Jason 2 satellite. The red crosses show the manoeuvre times,
           obtained independently from the TLE data. These plots should be contrasted with those in
           \figRef{fig:pitfalls:high_frequency_inclusion}, which show the differences in Keplerian elements. Observe
           how, for none of the three directions plotted, do the large propagation errors align particularly well with
           the manoeuvre timestamps.}
       \label{fig:pitfalls:high_frequency_positions}
\end{figure}

\begin{figure}[ht!]
       \centering
       \begin{subfigure}[]{0.45\linewidth}
           \includegraphics[width=\linewidth]{figs/osculating_to_osculating_a_diff_offset1.pdf}
           \caption{Difference in osculating semi-major axes}
           \label{fig:pitfalls:high_frequency_inclusion:osculating_a}
       \end{subfigure}
       \begin{subfigure}[]{0.45\linewidth}
           \includegraphics[width=\linewidth]{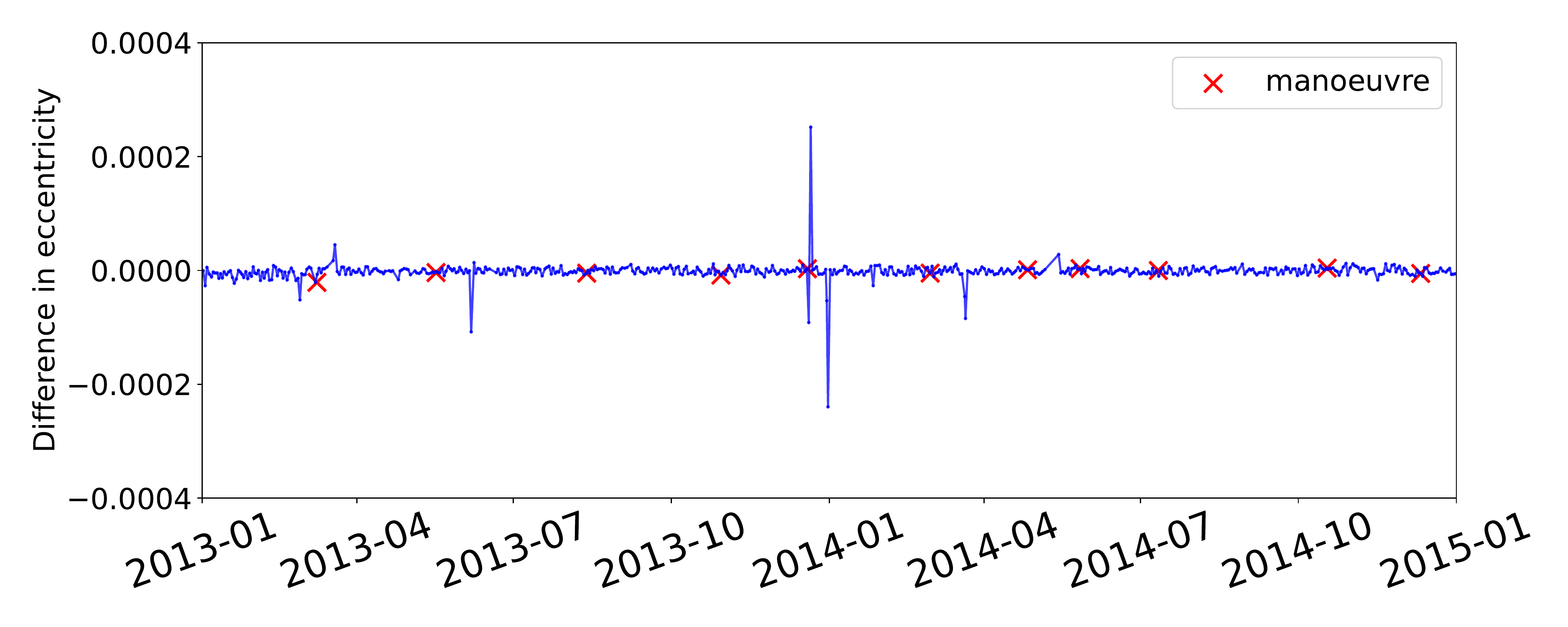}
           \caption{Difference in osculating eccentricities}
           \label{fig:pitfalls:high_frequency_inclusion:osculating_e}
       \end{subfigure}
       \begin{subfigure}[]{0.45\linewidth}
           \includegraphics[width=\linewidth]{figs/mean_to_mean_SMA_diff_offset1.pdf}
           \caption{Difference in mean semi-major axes}
           \label{fig:pitfalls:high_frequency_inclusion:mean_a}
       \end{subfigure}
       \begin{subfigure}[]{0.45\linewidth}
           \includegraphics[width=\linewidth]{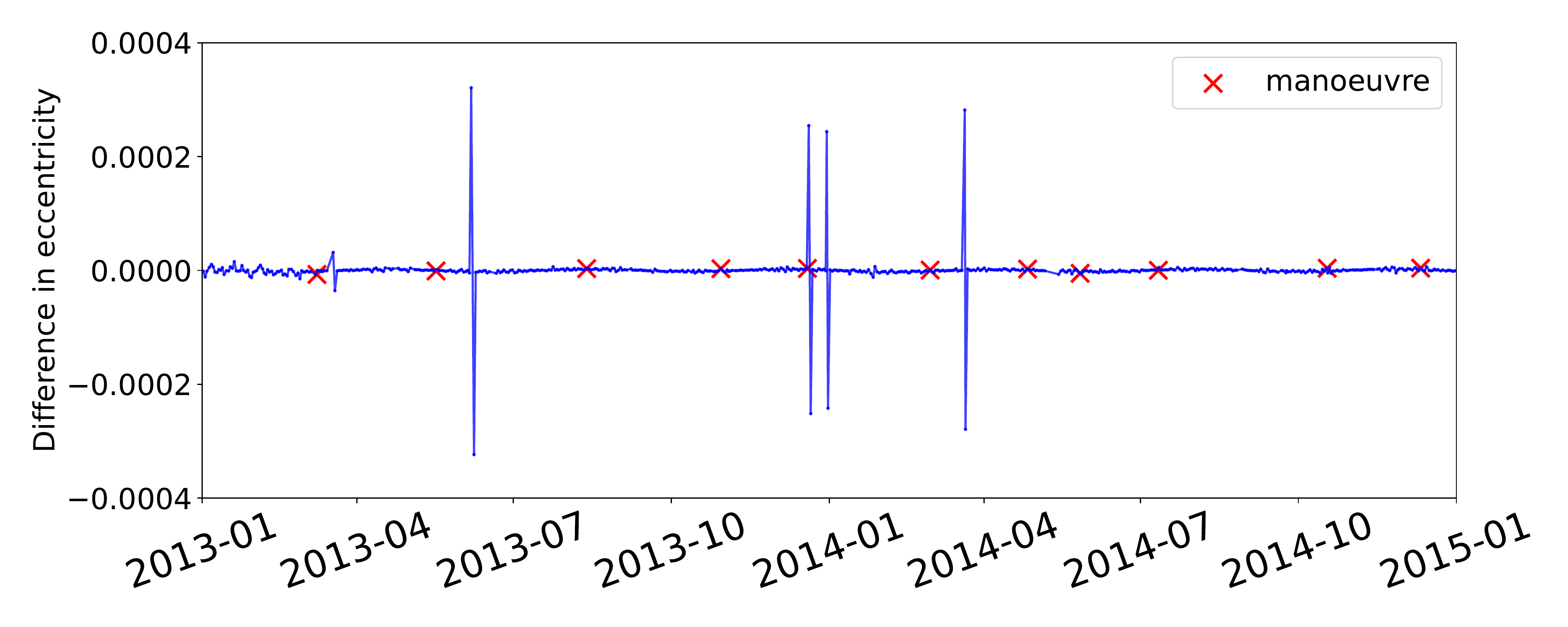}
           \caption{Difference in mean eccentricities}
           \label{fig:pitfalls:high_frequency_inclusion:mean_e}
       \end{subfigure}
       \begin{subfigure}[]{0.45\linewidth}
           \includegraphics[width=\linewidth]{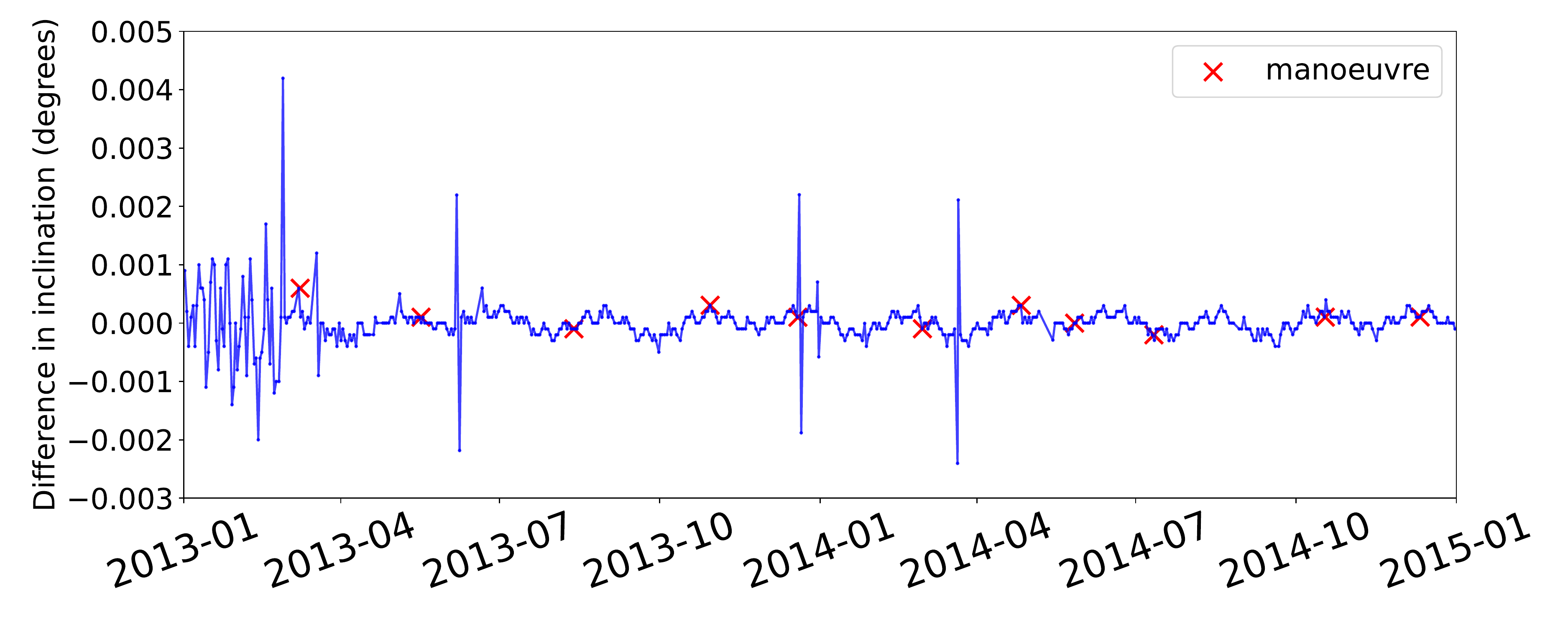}
           \caption{Difference in osculating inclinations}
           \label{fig:pitfalls:high_frequency_inclusion:osculating_i}
       \end{subfigure} \\
       \begin{subfigure}[]{0.45\linewidth}
           \includegraphics[width=\linewidth]{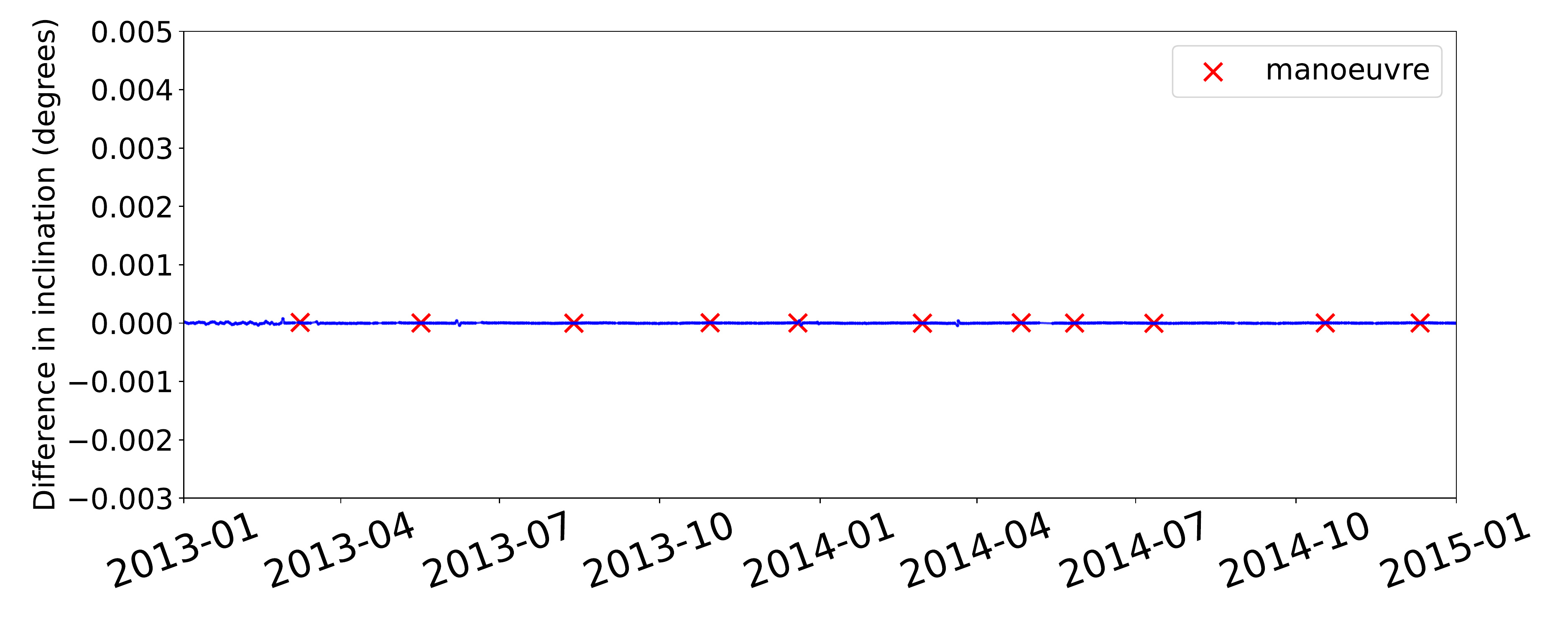}
           \caption{Difference in mean inclinations}
           \label{fig:pitfalls:high_frequency_inclusion:mean_i}
       \end{subfigure}
       \caption{\footnotesize
           Compares different propagations of TLE data to the observed orbital elements at the subsequent epoch.
           (\protect\subref{fig:pitfalls:high_frequency_inclusion:osculating_a}),
           (\protect\subref{fig:pitfalls:high_frequency_inclusion:osculating_e}) and
           (\protect\subref{fig:pitfalls:high_frequency_inclusion:osculating_i}) plot the results of using the full
           SGP4 algorithm (including non-Keplerian terms) to propagate the TLE from each epoch to the subsequent
           epoch. The SGP4 algorithm is also used to add the high frequency terms at the subsequent epoch in order to
           arrive at osculating Keplerian elements. The difference between the propagated and non-propagated elements
           is then found.
           (\protect\subref{fig:pitfalls:high_frequency_inclusion:mean_a}),
           (\protect\subref{fig:pitfalls:high_frequency_inclusion:mean_e}) and
           (\protect\subref{fig:pitfalls:high_frequency_inclusion:mean_i}), by contrast, show the results of
           propagating TLEs using only the low-frequency terms of SGP4, resulting in propagated mean elements. These
           elements are then compared directly against the TLE elements in the subsequent epoch. The plots show this
           analysis for the Jason 2 satellite.  The red crosses show the manoeuvre times, obtained independently from
           the TLE data. A number of differences between these approaches can be observed. The propagation errors for
           the inclination contain substantially less noise when this is done using the mean elements. On the other
           hand, the computation of these errors for the eccentricity appears quite unstable.
       }
       \label{fig:pitfalls:high_frequency_inclusion}
\end{figure}

\begin{figure}[ht!]
       \centering
       \begin{subfigure}[]{0.45\linewidth}
           \includegraphics[width=\linewidth]{figs/osculating_to_osculating_a_diff_offset5.pdf}
           \caption{Difference in osculating semi-major axes}
           \label{fig:pitfalls:high_frequency_inclusion:osculating_a_offset_5}
       \end{subfigure}
       \begin{subfigure}[]{0.45\linewidth}
           \includegraphics[width=\linewidth]{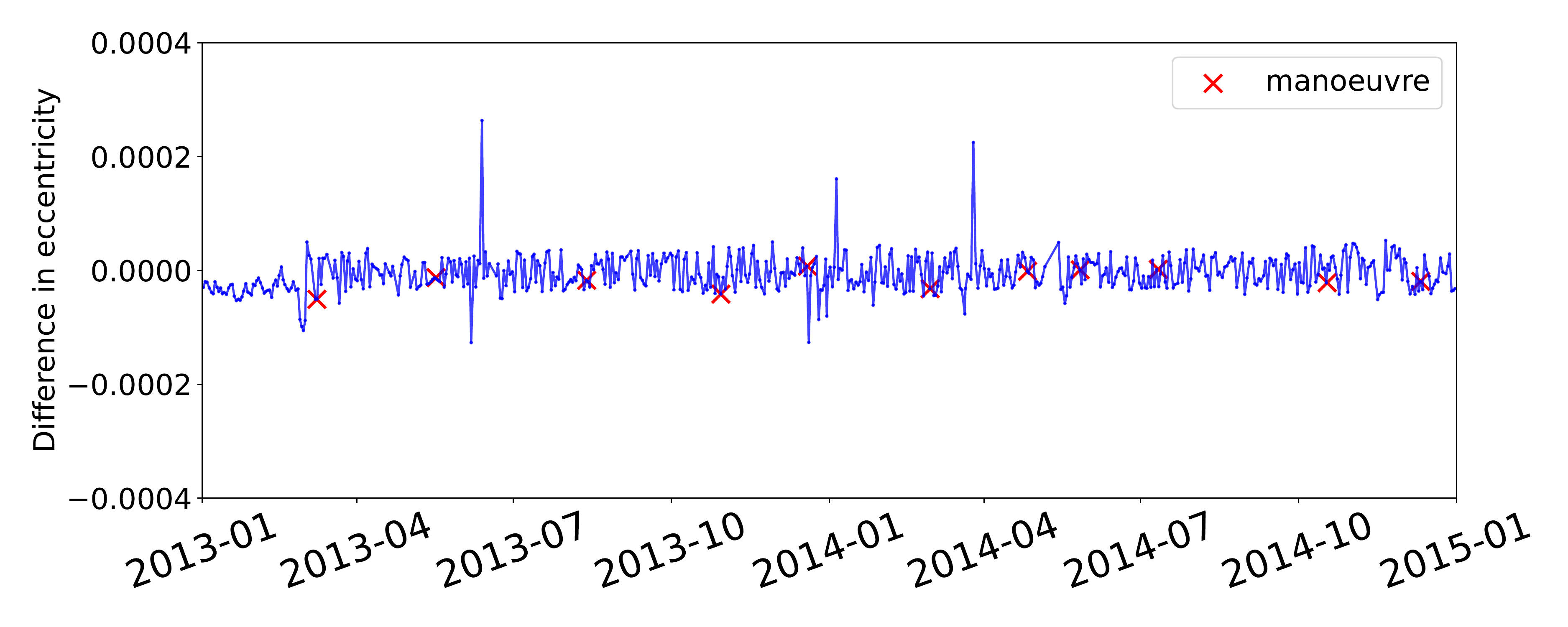}
           \caption{Difference in osculating eccentricities}
           \label{fig:pitfalls:high_frequency_inclusion:osculating_e_offset_5}
       \end{subfigure}
       \begin{subfigure}[]{0.45\linewidth}
           \includegraphics[width=\linewidth]{figs/mean_to_mean_SMA_diff_offset5.pdf}
           \caption{Difference in mean semi-major axes}
           \label{fig:pitfalls:high_frequency_inclusion:mean_a_offset_5}
       \end{subfigure}
       \begin{subfigure}[]{0.45\linewidth}
           \includegraphics[width=\linewidth]{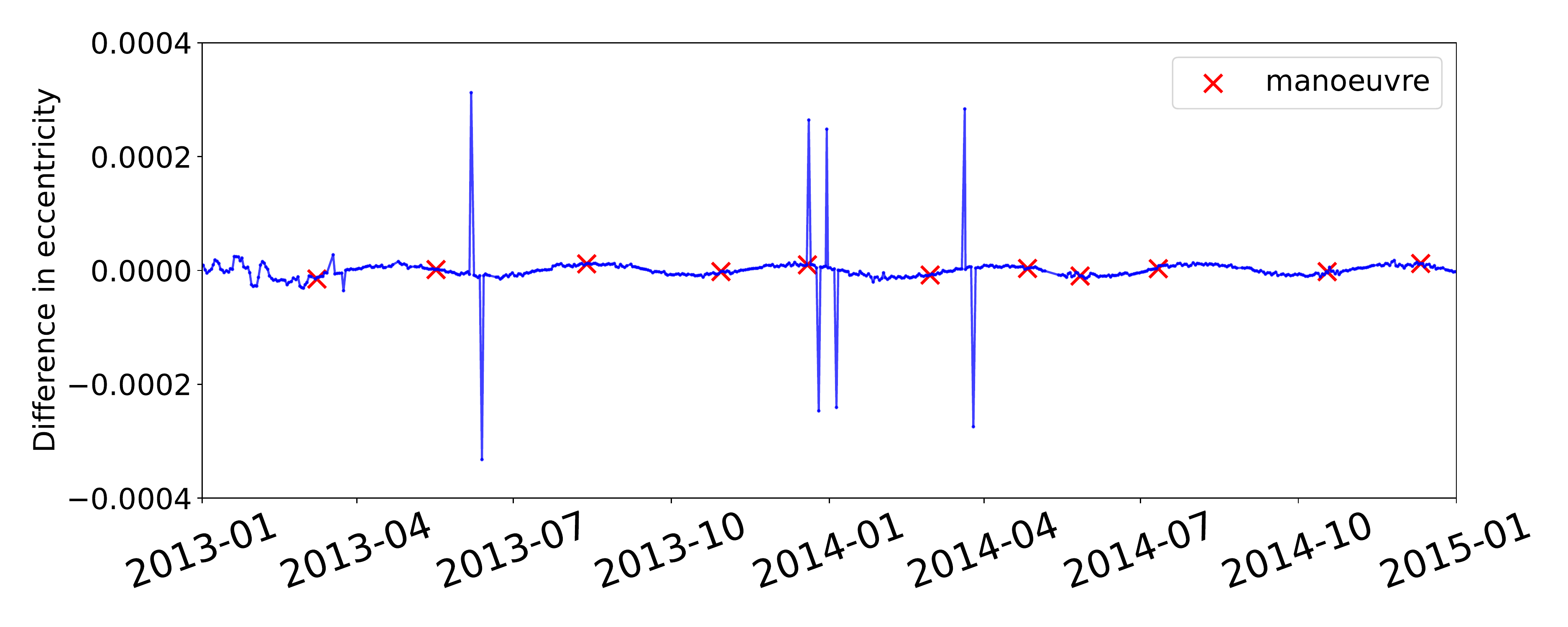}
           \caption{Difference in mean eccentricities}
           \label{fig:pitfalls:high_frequency_inclusion:mean_e_offset_5}
       \end{subfigure}
       \begin{subfigure}[]{0.45\linewidth}
           \includegraphics[width=\linewidth]{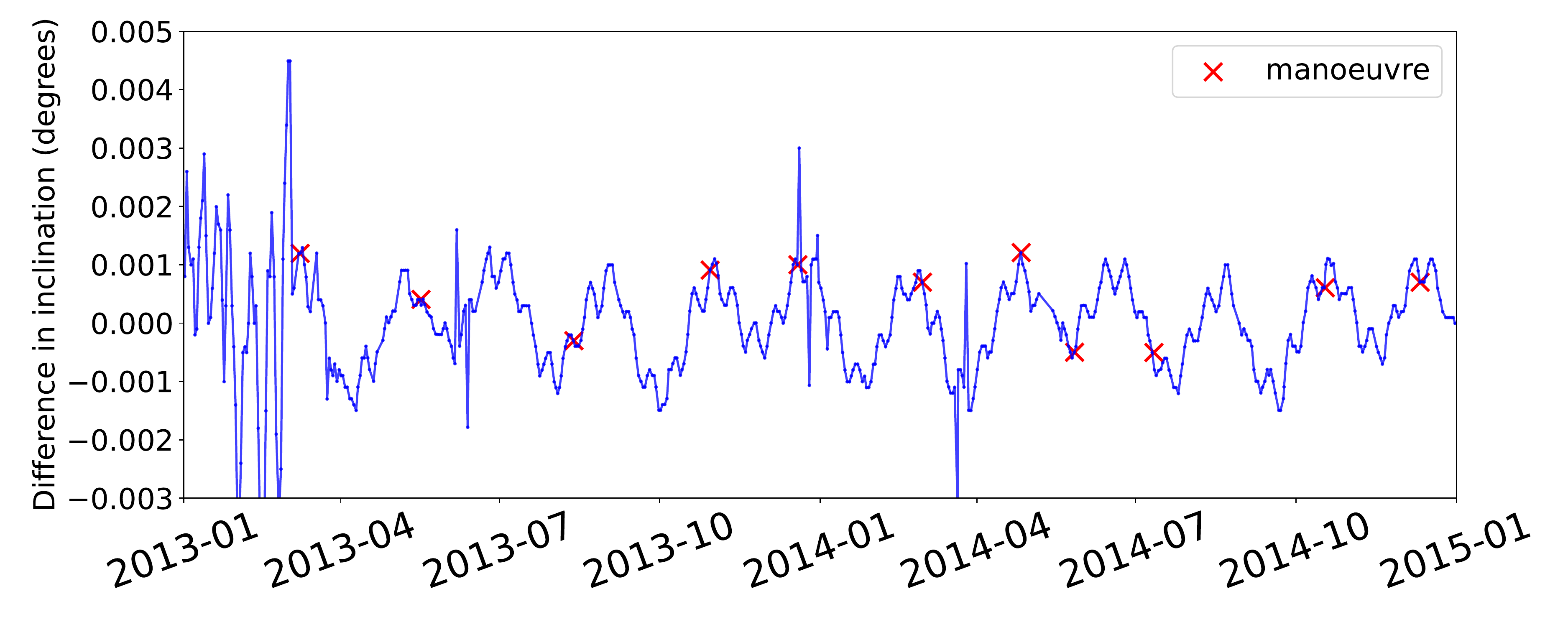}
           \caption{Difference in osculating inclinations}
           \label{fig:pitfalls:high_frequency_inclusion:osculating_i_offset_5}
       \end{subfigure} \\
       \begin{subfigure}[]{0.45\linewidth}
           \includegraphics[width=\linewidth]{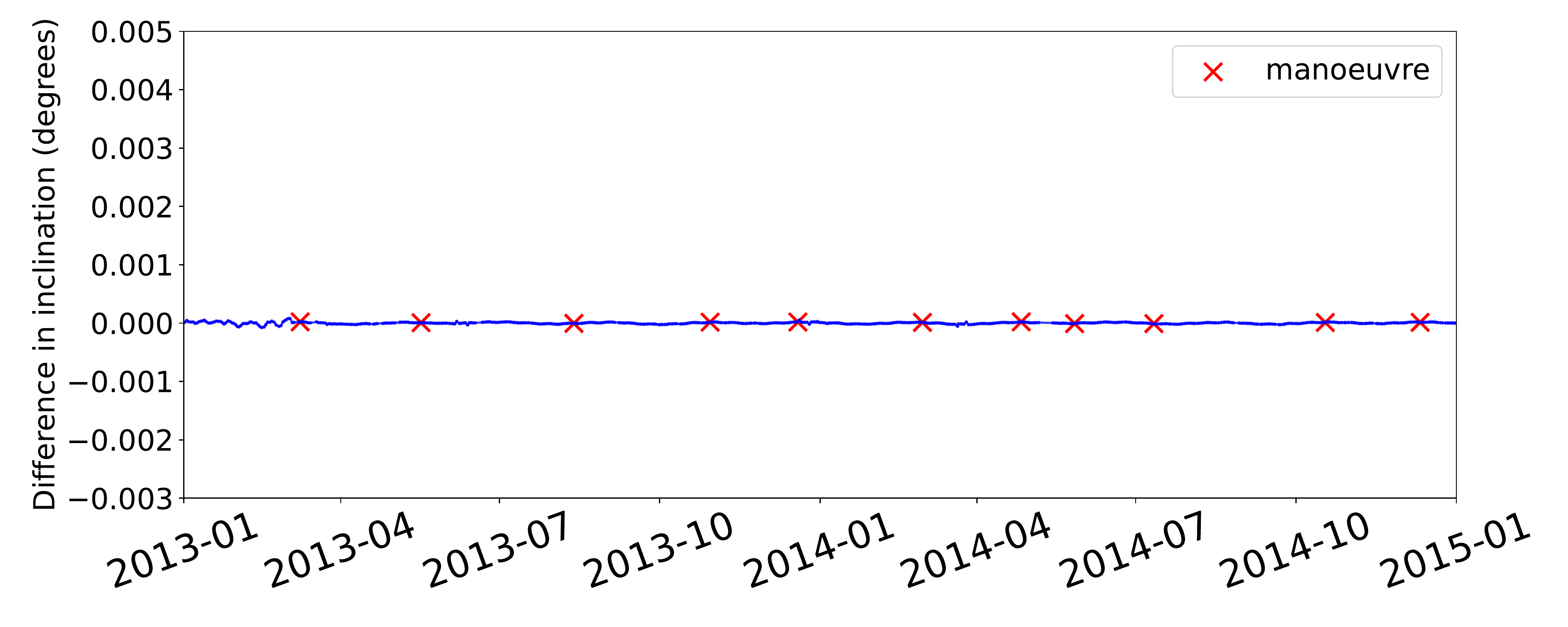}
           \caption{Difference in mean inclinations}
           \label{fig:pitfalls:high_frequency_inclusion:mean_i_offset_5}
       \end{subfigure}
       \caption{
           Compares different propagations of TLE data to the observed orbital elements five epochs later.
           Contains the same plots as \figRef{fig:pitfalls:high_frequency_inclusion}, but the propagation is
           performed over 5 epochs as opposed to 1. We observe significantly more noise when the propagations are
           performed using the full SGP4/SDP4 model (that is, by comparing osculating elements), as compared to only
           propagating the mean elements.
       }
       \label{fig:pitfalls:high_frequency_inclusion_offset_5}
\end{figure}

This appendix contains extra plots demonstrating the first pitfall, as discussed in \secRef{sec:pitfalls:full_sgp4}.
They were excluded from that section for the purpose of brevity. Whereas
\figRef{fig:pitfalls:propagation_type_comparison} shows plots of propagation residuals in the radial direction and
along the semi-major axis, \figRef{fig:pitfalls:high_frequency_positions} includes residuals in the in-track and
cross-track directions. Similarly, \figRef{fig:pitfalls:high_frequency_inclusion} and
\figRef{fig:pitfalls:high_frequency_inclusion_offset_5} include residuals in the eccentricity and inclination.

\end{appendices} 
\FloatBarrier

\bibliography{SatelliteTLEs}
\bibliographystyle{plain}

\end{document}